%% file: Manuscript.tex
\documentclass[unnumsec,webpdf,modern,large]{oup-authoring-template}%

\input{definitions}

\usepackage{graphicx}
\usepackage{xcolor}
\usepackage{array}
\usepackage{tabularx}
\usepackage{url}
\usepackage{pdfpages}

\begin{document}

\journaltitle{}
\DOI{To appear in PNAS Nexus (DOI: \href{https://doi.org/10.1093/pnasnexus/pgag345}{10.1093/pnasnexus/pgag345})}
\copyrightyear{}
\pubyear{}


\title[How Does Hard Magnetic Soft Matter Deform?]{How Does Hard Magnetic Soft Matter Deform?}

\author[1,2]{Daniel Katusele}
\author[3]{Carmel Majidi}
\author[4]{Liping Liu}
\author[5]{Pradeep Sharma}
\author[2,3,6,$\ast$]{Kaushik Dayal}

\address[1]{Program in Computational Mechanics, \orgname{Carnegie Mellon University}}
\address[2]{Department of Civil and Environmental Engineering, \orgname{Carnegie Mellon University}}
\address[3]{Department of Mechanical Engineering, \orgname{Carnegie Mellon University}}
\address[4]{Department of Mathematics, \orgname{Rutgers University}}
\address[5]{Department of Mechanical Engineering, \orgname{University of Houston}}
\address[6]{Center for Nonlinear Analysis, \orgname{Carnegie Mellon University}}

\corresp[$\ast$]{To whom correspondence should be addressed: \href{email:Kaushik.Dayal@cmu.edu}{Kaushik.Dayal@cmu.edu}}

\abstract{
    Hard Magnetic Soft Materials, i.e., elastomers seeded with hard magnetic particles, convert magnetic energy into complex motion and have become central to the design of untethered soft robots, biomedical devices, and, in general, for enabling fast, wireless actuation. 
    Their behavior depends critically on how the surrounding soft matrix deforms and transmits torque to each magnetized inclusion, yet there is no consensus on the correct mapping between macroscopic strain and microscopic motion. 
    Competing models predict conflicting particle rotations and magnetic responses under identical loads. 
    Here, we construct counterexamples that show the key shortcomings in these models, and then establish a first-principles framework that leads to an optimal deformation mapping. 
    The resulting closed-form law unifies conflicting models and shows that they are, in fact, simply different limits of our general approach. Our work permits the rational design of magnetically actuated soft devices and a unified view of deformation in magnetic soft matter, and broadly, in similar architectured materials.
}


\otherabstract[Significance statement]{
    Hard-magnetic soft materials, which couple elastic deformation with remnant magnetization, are central to the design of wireless soft robots, adaptive systems, and medical catheters.
    In all of these technologies, device-scale shape control depends on how the programmed magnetization evolves under deformation.
    Yet, there is controversy regarding the fundamental micromechanisms underpinning their multiphysical coupling.
    This work resolves that debate by deriving an analytic, energy-based law that reconciles previously competing hypotheses on the deformation mappings. 
    The resulting closed-form expressions predict the motion of magnetic particles across a broad range of geometries and loading conditions, and recover all known limits; specifically, it shows that seemingly conflicting models are simply different limits of our general approach.
    Beyond clarifying the mechanics of magnetic soft matter, the framework provides a systematic and predictive foundation for designing efficient, programmable actuation in emerging soft robotic and multifunctional materials.
}

\maketitle

\begin{figure}[h!]
    \centering
    \includegraphics[width=0.93\columnwidth]{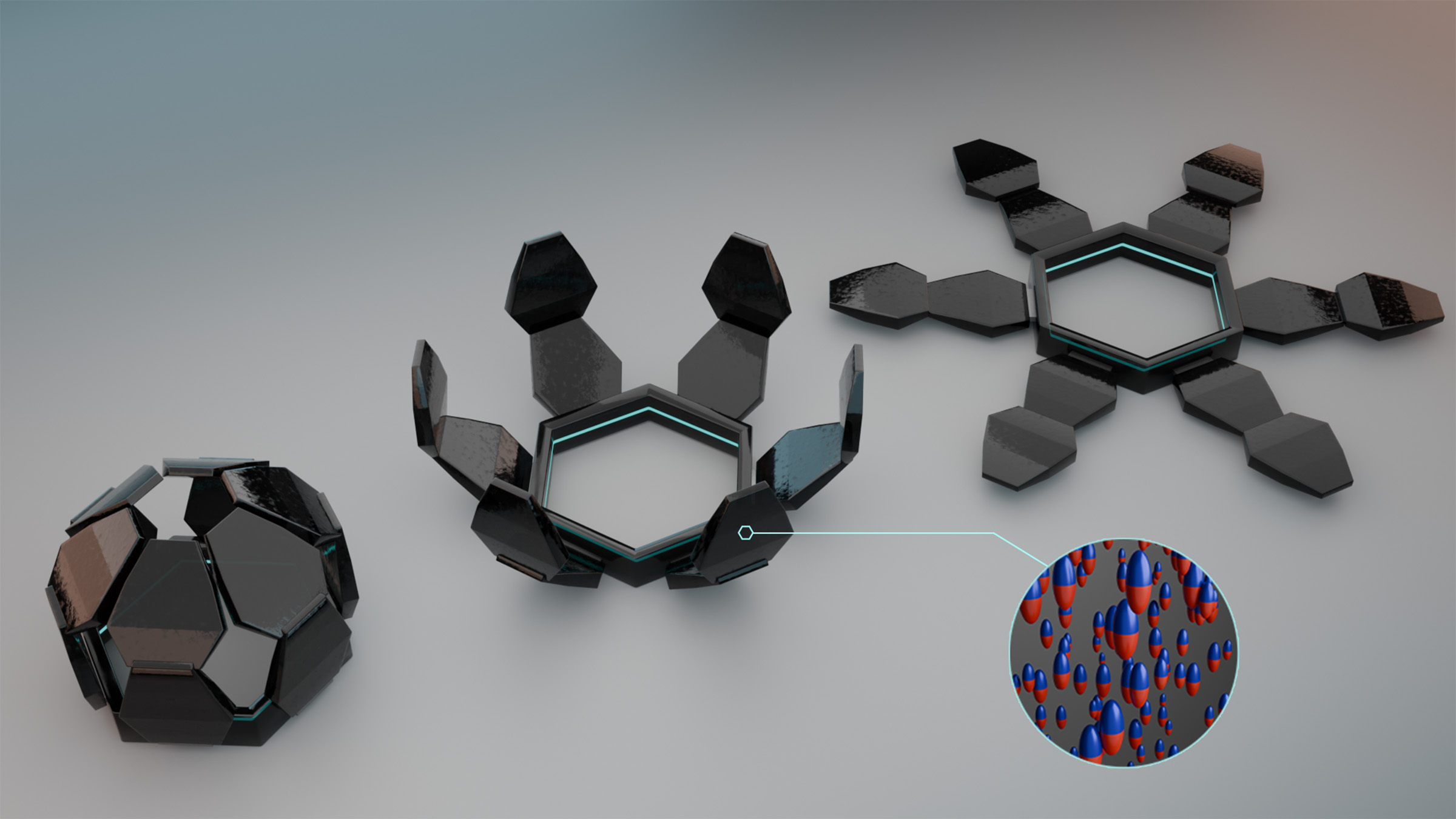}
    \caption{Schematic of a hard magnetic soft robotic device. Programmed magnetic particle distributions enable precisely controlled opening and closing motions in response to an external magnetic field.}
    \label{fig:1}
\end{figure}

Soft materials capable of remote, reversible transformation are transforming robotics, electronics, and medicine. Among them, hard-magnetic soft materials, i.e. elastomers filled with high-coercivity magnetic particles \cite{Bira2020MEreview,moreno2022hybrid}, stand out for combining extreme deformability with strong, programmable magnetic response. When exposed to an external magnetic field, the residual magnetization of each particle exerts a torque on the compliant matrix, driving rapid bending, twisting, and folding \cite{kim2022magnetic}. The ability to ``print'' patterns of magnetic dipoles within soft polymers has made possible untethered soft robots, adaptive lenses, and reconfigurable surfaces \cite{Kim2018PrintingFerromagnetic, Zhao2019MechanicsHardMagneticSoft,cui2019nanomagnetic,schumann2017characterisation,TANG2021106424,narayanan2024hard}. The pioneering work in reference \cite{Kim2018PrintingFerromagnetic} has inspired significant theoretical, experimental, and translational research across the gamut of fields ranging from solid mechanics \cite{Zhao2019Mechanics, Zhao2022Topology, li2022digital, Rahmati2023Theory}, additive manufacturing \cite{Wu2020Evolutionary, Wu2020Multimaterial, Wang2025Direct}, robotic devices \cite{Zhao2021MagneticSoftRobots, Hu2018SmallScale, Rahmati2023Giant}, biomedicine \cite{Chang2025MilliSpinner, Sitti2020Magnetic}, among others.
For instance, in shape-forming catheters, the programmed magnetization profiles must be designed to steer soft slender devices through highly constrained anatomical pathways \cite{Lloyd2020OptimalDesign,Pittiglio2022PatientSpecific,Davy2023Computational}; because device-level models cannot resolve every magnetic particle, they require a robust and physics-based law connecting deformation, particle rotation, and magnetization.

These materials bridge two scales of physics: the magnetic torque on micron-sized inclusions and the resulting macroscopic deformation. 
Their key attribute that there is remnant or spontaneous magnetization that persists without an applied external field distinguishes hard-magnetic materials (e.g., NdFeB) from the soft-magnetic fillers used in earlier magnetorheological elastomers \cite{Wu_2020Multifunctional}. 
In hard-magnetic systems, magnetization persists without continuous field application, enabling large, fast, and autonomous motions (Fig. 1) \cite{hu2018small,Li2025Magneticfield}.

Despite rapid experimental progress, there is controversy on the fundamental micro-mechanism of deformation, stemming from the manner in which the embedded rigid magnetic particles rotate within the deforming matrix. 
This rotation controls not only the global shape change but also the local magnetic torque and mechanical stress distribution. \par
In early works on the modeling of such materials, Zhao et al. \cite{Zhao2019MechanicsHardMagneticSoft} used the material mapping model for magnetic flux (and magnetization) of the form
\begin{align}
\label{eqn:zhao-map}
    \bfB_0 = J \bfF^{-1} \bfb
\end{align}
in which $\bfB_0$ and $\bfb$ are the magnetic flux in the reference and deformed configurations, respectively, and $\bfF$ and $J$ are the macroscopic deformation gradient and its determinant; this mapping is developed from Gauss' law for magnetism in conjunction with Nanson's transformation for areas under deformation.  
Here $\bfF=\nabla_{\bfx}\bfy$ denotes the deformation gradient of the deformation map from the reference to the current configuration, and $J=\det\bfF$ is the corresponding Jacobian determinant.
The magnetization, under the assumption of linearity with magnetization proportional to magnetic flux through the inverse of the magnetic permeability, will follow the same transformation. 
This model uses a reference (undeformed) configuration and a push-forward transformation to compute the magnetic response in the deformed configuration. 
An experimental test under pure bending with negligible stretching validates the model \cite{Zhao2019MechanicsHardMagneticSoft}, leading to its adoption in later works by many others \cite{LU2024104874}. 
A key consequence of this model is that the magnetic flux (and magnetization) changes with mechanical stretching through the deformation gradient as in \eqref{eqn:zhao-map}. 
Therefore, in the absence of external/applied magnetic field, the residual magnetic flux (and magnetization) will depend on stretch.

Recent work \cite{danas2024stretch} has questioned the validity of the material mapping assumption, arguing that it can produce nonphysical predictions when finite stretches are present. 
The primary point of concern is that, in a finite-strain framework, the magnetic field must be defined in the deformed configuration \cite{KANKANALA20042869}, and therefore lacks a physical significance when expressed in the reference configuration. 
Moreover, several studies, including those by Yan et al. and Danas et al. \cite{YAN2023105095, DANAS2012120}, together with supporting numerical and experimental results \cite{MUKHERJEE2021104361}, suggest that the remanent magnetic flux may not stretch -- or transform proportionally -- with tensile or compressive load. Instead, these works indicate that a primarily rotational, rather than stretch-dependent, evolution of magnetization can occur. 
Such behavior is consistent with the expectation that rigid magnetic inclusions are largely unaffected by the deformation of the surrounding soft matrix.\par

To address the shortcomings of the material mapping, several authors have proposed mapping the magnetization through a rotation rather than the full deformation gradient \cite{danas2024stretch, YAN2023105095, Zhou2025Theory}. 
In these approaches, the rotation corresponding to the polar decomposition of the deformation gradient is taken as the operative transformation of the magnetization in finite deformation analyses \cite{danas2024stretch, YAN2023105095}. 
Under this assumption, only the direction of the magnetization changes with deformation, while its magnitude remains constant. This modification allegedly addresses the main flaw in the material mapping model \cite{Zhao2019MechanicsHardMagneticSoft}. The formulation by Danas and Reis \cite{danas2024stretch}, for example, introduces an internal variable $\mathcal{H}^r$ defined in a stretch-free intermediate configuration such that
\begin{align}
    \bfm = \bfR^{\text{PD}} \mathcal{H}^r
\end{align}
where $\bfm$ denotes the magnetization in the deformed configuration and $\bfR^{\text{PD}}$ is the rotational component of the polar decomposition of $\bfF$. This assumption is intended to capture the idea that, in the absence of stretch effects, the remanent magnetization evolves only through rotation. The internal variable introduced in this formulation is therefore related to the deformed magnetization through $\bfR^{\text{PD}}$. An alternative treatment proposed by Yan et al. \cite{YAN2023105095} also employs the rotational part of the deformation gradient, but does so within a push-forward mapping of the magnetization from the reference to the deformed configuration. In that model, the transformation of magnetization is expressed as
\begin{align}
    \bfm = J^{-1} \bfR^{\text{PD}} \bfM_0.
\end{align}

Table~\ref{tab:mapping_comparison} summarizes these two existing mappings alongside the MIA mapping developed below, highlighting the distinct assumptions and limiting behavior that organize the remainder of the paper.

\begin{table*}[htb!]
\centering
\small
\setlength{\tabcolsep}{3.5pt}
\renewcommand{\arraystretch}{1.18}
\caption{Side-by-side comparison of the three deformation-to-magnetization mappings discussed in this work. The material mapping uses the full deformation gradient, the polar-decomposition mapping uses only the macroscopic polar rotation, and the MIA mapping uses the rigid-body rotation of the inclusion obtained from the micromechanical variational problem.}
\label{tab:mapping_comparison}
\begin{tabularx}{\textwidth}{|>{\raggedright\arraybackslash}p{0.16\textwidth}|>{\raggedright\arraybackslash}X|>{\raggedright\arraybackslash}X|>{\raggedright\arraybackslash}X|}
\hline
 & Material mapping & Polar-decomposition mapping & MIA mapping \\
\hline
Representative law & $\bfm \propto J^{-1}\barF\bfM_0$ & $\bfm = \bfR^{\text{PD}}\bfM_0$ & $\bfm=\bfR^{\text{MIA}}\bfM_0$ \\
\hline
Main assumption & Magnetization is convected by the macroscopic deformation & Magnetization follows the macroscopic polar rotation only & Magnetization obtained from micromechanics \\
\hline
Preserves remanent-magnetization magnitude & Not in general, unless normalized & Yes & Yes \\
\hline
Depends on inclusion shape/orientation & No & No & Yes, through $\eta$ and $\alpha$ \\
\hline
Pure stretch of an inclined anisotropic inclusion & Predicts stretch-dependent reorientation & Predicts no rotation & Predicts stretch- and geometry-dependent rotation \\
\hline
Recovered as a limit of MIA & Long, thin inclusion limit & Circular-inclusion/homogeneous-rotation limit & General finite-aspect-ratio result \\
\hline
\end{tabularx}
\end{table*}

\begin{figure}[htb!]
    \centering
    \includegraphics[width=1.0\columnwidth]{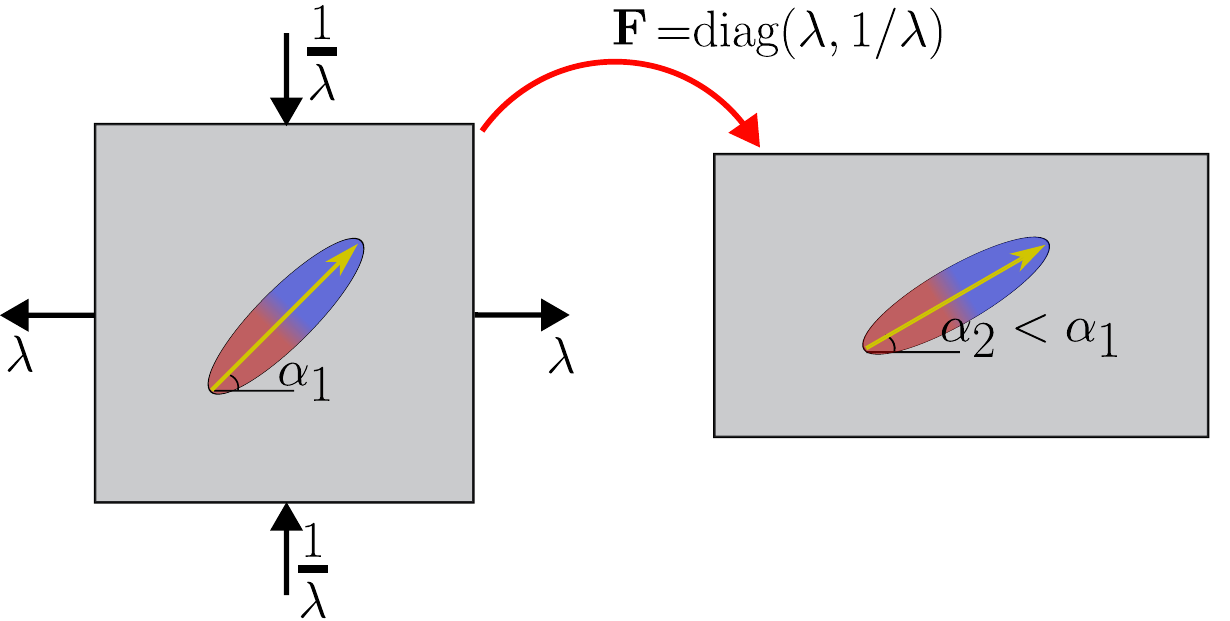}
    \caption{Counter-example to the polar-decomposition-based approaches for a generic inclined anisotropic particle. A single elliptic particle with initial orientation $\alpha_1$ subjected to horizontal elongation and vertical compression tends to rotate to align the long axis with the extensional direction, so that the deformed orientation satisfies $\alpha_2<\alpha_1$ for the representative case shown. However, the polar decomposition of this pure stretch is the identity, so all \emph{polar-decomposition-based mappings predict absolutely no rotation} for every initial angle.}
    \label{fig:counterexample}
\end{figure}

The formulations proposed as remedies to the material mapping model, unfortunately, themselves suffer from (hitherto unrecognized) significant conceptual and physical limitations. 
They all ultimately rely on the assumption that the rotational component of the polar decomposition governs how remanent magnetization transforms under deformation. 
However, this  assumption is highly questionable; it has been verified only in significantly limited contexts, mainly numerical studies of circular inclusions \cite{MUKHERJEE2022111513}, where shape anisotropy and torque imbalance are absent. 
In one such test \cite{MUKHERJEE2021104361}, the initial magnetization was aligned with the principal direction of stretch, resulting in zero average rotation; a result trivially reproduced by polar-decomposition-based models \cite{danas2024stretch, YAN2023105095}. 
Closer inspection reveals that this outcome stems from the models' construction rather than physical fidelity: whenever pure tension or compression is applied without shear, these formulations necessarily predict zero rotation. 
A simple counterexample exposes this deficiency clearly: an elliptical inclusion at a generic inclined angle to the tensile axis in a dilute composite rotates with increasing stretch, as illustrated in \figref{fig:counterexample}. The effect vanishes only in symmetry cases, such as a circular inclusion, no stretch, or alignment with a principal stretch direction. This example underscores a central flaw in using the polar rotation to map remanent magnetization: it cannot capture geometry-induced reorientation, even in the simplest mechanical loadings.

This leads to the fundamental unresolved question at the heart of how hard magnetic soft matter really deforms: {\bf how does the remanent magnetization transform under deformation?}
In mathematical terms, what is the constitutive mapping\footnote{Although, superficially, this appears to be a kinematic choice, it is in fact a material model with profound influence on the emergent mapping for a heterogeneous material. The ``correct'' model would consider the complete micro-geometry for every individual specimen, but this is neither feasible nor useful.} that relates the remanent magnetization to the deformed configuration: preserving its magnitude under finite stretch while accurately capturing rotation across the full spectrum of deformations characteristic of magnetic soft matter?\par

In this work, we answer this question through a first-principles micromechanical framework that constructs an energetic upper bound to the effective response of the composite. 
This variational approach yields an explicit, closed-form expression for the mapping of magnetization under finite deformation. 
The predictions of our framework are systematically compared with those of the material mapping \cite{Zhao2019MechanicsHardMagneticSoft} and the rotational mapping based on the polar decomposition \cite{danas2024stretch}, as well as with full-field numerical simulations, to test the accuracy and consistency of the proposed physics-based formulation.

A key outcome is our finding that the seemingly-conflicting prior models are simply different limit cases of our general first-principles approach.

\section{General Nonlinear Variational Framework} \label{sec:upper}

We begin by providing the general nonlinear energetic framework; specifically, the macroscopic energy is posed as a variational problem over admissible fields, enabling the construction of a computable upper bound.
We introduce a trial deformation that interpolates in an admissible way between the imposed homogeneous boundary deformation and a rigid-body motion of the inclusion, and then obtain the rotation of the inclusion by minimizing this bound over the free parameters related to the rotation.

\begin{figure}[htb!]
  \centering
  {\includegraphics[width=0.4\textwidth]{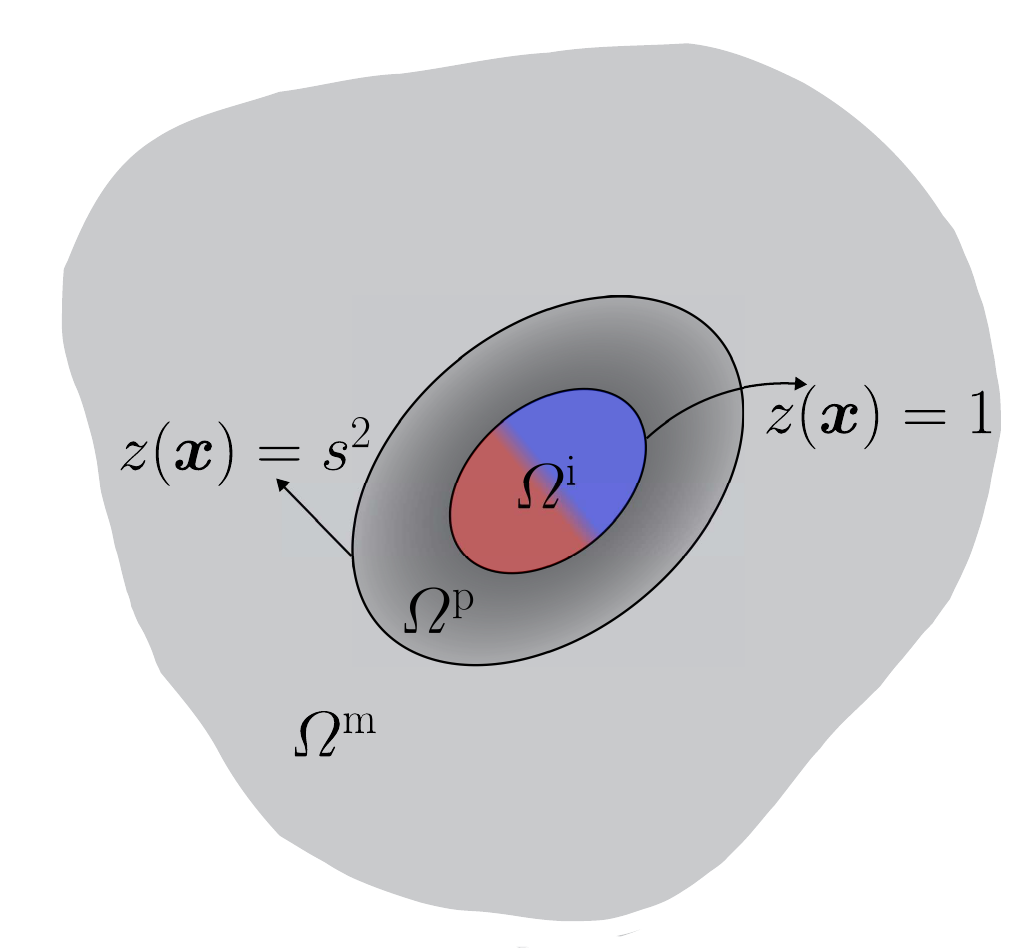}}
  \caption{Schematic of the level-set interpolation around the inclusion. The blue/red region $\OmInc$ represents the magnetic inclusion while the light blue $\OmI$ is the interpolation region. The region outside this, $\OmM$ is the matrix deforming as $\bfy = \barF \bfx$.}
  \label{fig:interpolation_domain}
\end{figure}

For simplicity and corresponding to the physical problem, we model the matrix as a near-incompressible Neo-Hookean soft solid and the inclusion as rigid.
Our ansatz for the deformation interpolates between the (given) homogeneous boundary condition $\bfy = \barF \bfx$ on the entire boundary and the rigid rotation and translation of the inclusion $\bfy = \bfR_0 \bfx + \bfc_0$, where $\bfR_0$ and $\bfc_0$ are to be determined by minimization.
We define $\overline{\bfF} := \langle \bfF \rangle$ as the average deformation gradient following \cite{hill1972constitutive}. 
The interpolation region, shown in  \figref{fig:interpolation_domain}, is parametrized by a level set function $z(\bfx)$ that characterizes the shape of the inclusion as well as the interpolation region around the inclusion.  
It divides the domain into three subdomains: the inclusion $\OmInc$ with $z(\bfx) \leq 1$; the interpolation region  $\OmI$ with $1 < z(\bfx) < s^2$; and the matrix $\OmM$ with $z(\bfx) \geq s^2$.
The scalar $s>1$ sets the size of the outer level set relative to the inclusion boundary. In other words, the inclusion boundary is $z=1$, the outer edge of the interpolation layer is $z=s^2$, and for an elliptical level set the semi-axes of the outer boundary are scaled by the factor $s$ relative to the inclusion semi-axes. Thus $s$ parameterizes the trial-field transition zone between rigid inclusion motion and affine far-field deformation; it is not introduced as a fitted material parameter. For the Neo-Hookean matrix considered below, the stationarity condition determining the optimal inclusion rotation is independent of $s$, although the upper-bound energy value can depend on the interpolation-region size.

The deformation map is formulated in terms of $z(\bfx)$ by interpolating between the far-field conditions $\bar{\bfF}$ and the rigid motion of the inclusion to get:
\begin{equation}
  \tbfy(\bfx) = \begin{cases} 
      \bfR_0 \bfx + \bfc_0 &  \bfx \in \OmInc 
      \\[1ex]
      \dfrac{\left(\bfR_0 \bfx + \bfc_0 \right) \left( z(\bfx) - s^2 \right) - \barF \bfx \left( z(\bfx) - 1\right)}{1-s^2}  &  \bfx \in \OmI  
      \\[1ex]
      \barF \bfx  &  \bfx \in \OmM.  
   \end{cases}  
   \label{eq:ansatz}
\end{equation}  
The corresponding deformation gradient $\tF = \nabla_{\bfx} \tbfy$ for this ansatz, specifically in the interpolation region ($\bfx \in \OmI$), is   
\begin{align}
    \tF
    =
    \frac{(\bfR_0 \bfx + \bfc_0 - \barF \bfx  ) \otimes \nabla_{\bfx} z  + \bfR_0 (z - s^2) - \barF (z - 1)}{1-s^2}
    \label{trial_defgradient}
\end{align}
with $\tF = \bfR_0$ and $\tF = \barF$ when $\bfx \in \OmInc$ and $\bfx \in \OmM$ respectively.

\subsection{Variational homogenization}
The stored energy density for the incompressible (i.e, $\det(\bfF) = 1$) neo-Hookean matrix is given by
\begin{align}
    W_{\text{matrix}}(\bfF)  = \frac{\mu}{2}  \trace \left(\bfF^T \bfF - \bfI \right) 
\end{align}
The stored energy in the rigid inclusion has the form:
\begin{align}
    W_{\text{inclusion}}(\bfF)  = \begin{cases} 
      0 & \text{if } \bfF \in \mathrm{SO}(2) \\[1ex]
      + \infty  & \text{otherwise }   \end{cases}
\end{align}
where $\mathrm{SO}(2)$ is the set of all rotations. 

The stored energy density for this two-phase composite is then:
\begin{align}
    W(\bfx, \bfF) =  (1 - \chi (\bfx)) W_{\text{matrix}} (\bfF)+ \chi(\bfx) W_{\text{inclusion}} (\bfF)
    \label{micro_energy}
\end{align}
where the characteristic function $\chi(\bfx)$ has a value of 0 or 1 if the material point $\bfx$ is the matrix or the inclusion respectively. 

The volume fraction of the inclusion is defined as the volume average $\phi := \langle \chi \rangle_{\Omega}$ over the entire domain. The macroscopic stored energy will be the infimum of the average stored energy \cite{hill1972constitutive,SIBONI2012205,Ca2023soft}, and is given by
\begin{align}
  \overline{W}(\overline{\bfF}) =  \underset{\bfF \in \mathcal{K}(\overline{\bfF})}{\inf} \frac{1}{| \Omega |} \int_{\Omega} W(\bfx, \bfF) \label{total_energy}  
\end{align}
where $\mathcal{K}(\overline{\bfF})$ is the set of kinematically admissible deformations; that is, gradient fields with positive determinant, satisfying the boundary conditions $\overline{\bfF}$, and that are a uniform rotation in the inclusion.
The infimum allows for the formation of infinitely-fine microstructure due to non-convexity in the stored energy \cite{Ca2023soft}. 
However, we will restrict our attention to smooth fields and the macroscopic energy can be written as the minimum of the average stored microscopic energy in \eqref{micro_energy} in the form \cite{hill1972constitutive, SIBONI2012205,Ca2023soft}:
\begin{align}
    \overline{W}(\overline{\bfF}) =  (1 - \phi)\underset{\bfF \in \mathcal{K}(\overline{\bfF})}{\min} ~\frac{1}{| \Omega^{\text{el}} |} \int_{\Omega^{\text{el}}} W_{\text{matrix}}(\bfx, \bfF) \label{total__effective_energy}
\end{align}
where $\Omega^{\text{el}} = \OmI \cup \OmM$. 
The minimization in \eqref{total__effective_energy} is over the entire set of admissible deformation gradients. 
Consequently, the ansatz provides an upper bound for the macroscopic energy:
\begin{align}
     \overline{W}(\overline{\bfF}) \leq \overline{W}(\tF).
     \label{eq:variational_bound}
\end{align}
Several methods have been proposed for generating variational bounds to the energy, for example, by generating fictitious linear comparison composites among several approaches \cite{Ca2023soft}.
In contrast, our goal is to identify the kinematic features of the deformation corresponding to the rigid motion of the inclusion.
Therefore, our strategy is to use the ansatz in \eqref{trial_defgradient} as a trial deformation, and subsequently compute $\bfR_0$ and $\bfc_0$ by minimizing the energy of the upper bound defined above.

\section{Derivation of the Optimal Mapping} \label{sec:rotation}

We next specialize the framework to derive an closed-form rotation mapping for a single rigid elliptical inclusion in a two-dimensional plane-strain setting. This restriction is made to obtain a compact expression with the rotation described by a single angle. The variational construction in the previous section is more general; in three dimensions, an ellipsoidal inclusion can be described by an analogous quadratic level set and the rigid motion is obtained by minimizing the same upper-bound energy over $\bfR_0\in \mathrm{SO}(3)$, rather than over one planar angle. The explicit expression below is therefore the planar specialization of the general energetic idea. For a Neo-Hookean matrix, the resulting planar rotation depends on the macroscopic deformation and inclusion shape parameters but is independent of the interpolation lengthscale $s$.

The shape of the inclusion and initial orientation of the inclusion are described by the shape tensor $\bfP$ and rotation $\bfR_{\alpha}$ expressed as 
\begin{align}
    \bfP_{\eta} = \begin{pmatrix}
        \dfrac{1}{\eta^2} & 0 \\
        0 & 1
    \end{pmatrix} \quad  \text{and} \quad \bfR_{\alpha} = \begin{pmatrix}
        \cos{\alpha} & \sin{\alpha} \\
        -\sin{\alpha} & \cos{\alpha}
    \end{pmatrix}
\end{align}
in which $\eta$ is the aspect ratio of the inclusion with the major axis of the ellipse aligning with the $x$-axis, and $\alpha$ is the orientation of the inclusion in the stress-free configuration. To distinguish particle geometry from magnetization, we denote the geometric orientation by $\alpha_{\mathrm g}$ and retain $\alpha\equiv\alpha_{\mathrm g}$ in the historical mechanical formulas below. Independently, $\beta$ denotes the laboratory-frame direction of the initial remanent magnetization. The domains
\begin{equation}
\begin{split}
    \OmInc &= \{ \bfx : \bfx \cdot \bfR_{\alpha}^T \bfP_{\eta} \bfR_{\alpha} \bfx\leq 1 \}, \\  \OmI &= \{ \bfx : 1 < \bfx \cdot \bfR_{\alpha}^T \bfP_{\eta} \bfR_{\alpha} \bfx\leq s^2 \}, \\ \OmM &= \{ \bfx : \bfx \cdot \bfR_{\alpha}^T \bfP_{\eta} \bfR_{\alpha} \bfx > s^2 \}
    \label{eq:domain_decompose}
\end{split}
\end{equation}
define the inclusion, interpolation region, and far-field matrix.
For this ellipse, $s$ has the simple geometric meaning of a dilation factor: the interpolation-region outer boundary has the same orientation and aspect ratio as the inclusion but semi-axes multiplied by $s$. The minimization below is formally written over $s$ as part of the upper-bound construction, but the closed-form rotation obtained from the Neo-Hookean energy does not depend on the final value of $s$.
The stored energy representing the variational bound in \eqref{eq:variational_bound} can then be written as
\begin{equation}
    \overline{W}(\tF) 
    =  
    \underset{\bfR_0, \bfc_0, s}{\min}  \frac{1}{|\Omega|} 
    \left( 
    \begin{aligned}
        & 
        \int_{\OmI} \frac{\MUm}{2}  \trace \left(\tF^T \tF - \bfI \right)  
        \\ 
        & 
        + \int_{\OmM} \frac{\MUm}{2}  \trace \left(\bfF_0^T \bfF_0 - \bfI \right) 
    \end{aligned}
    \right) 
    \label{total_energy_min}
\end{equation}
in which we seek to minimize the energy with respect to the rotation, translation, and the width of the interpolation region. 
Notice that the contribution from the matrix in the far-field region is a constant (given $\bfF_0$), and hence does not appear in the minimization.
The definition of the form of the interpolation region in \eqref{eq:domain_decompose} simplifies the explicit computation.
For more complex shapes, it is possible to use different implicit or explicit functions in our overall approach.

\subsection{The General Mapping}

For the analysis that follows, we consider a macroscopic deformation gradient and a rotation given by:
\begin{align}
    \barF = \begin{pmatrix}
        F_{11} & F_{12} \\
        F_{21} & F_{22}
    \end{pmatrix}
    , \quad
    \bfR^{\Box} = \begin{pmatrix}
        \cos\theta^{\Box} & \sin\theta^{\Box} \\
        -\sin\theta^{\Box} & \cos\theta^{\Box}
    \end{pmatrix}
\end{align}
in which we have introduced the superscript $\Box$ to be defined for rotations obtained with different methods. 
The rotation $\bfR^{\text{MIA}}$, where ``MIA'' stands for Micromechanical Interpolation Approach, is computed from minimizing the upper bound \eqref{total_energy_min}. 
We have worked in polar coordinates for algebraic simplicity to enable explicit computation of the integration and derivatives, and assumed near-incompressibility, to finally obtain the general expression:
\begin{align}
\theta^{\text{MIA}}
  &= \tan^{-1}\left[
     \frac{%
     \begin{aligned}
         &(F_{11}-F_{22})(\eta^4-1)\sin(2\alpha) \\
         &\quad -(F_{12}+F_{21})(\eta^4-1)\cos(2\alpha) \\
         &\quad +(F_{12}-F_{21})(\eta^4+6\eta^2+1)
       \end{aligned}
       }{%
     \begin{aligned}
         &(F_{12}+F_{21})(\eta^4-1)\sin(2\alpha) \\
         &\quad +(F_{11}-F_{22})(\eta^4-1)\cos(2\alpha) \\
         &\quad +(F_{11}+F_{22})(\eta^4+6\eta^2+1)
       \end{aligned}
     }{       
     }
     \right]
     \label{eq:rotation_map}
\end{align}
The details of the derivation are in the Supplemental Material.
The final expression above depends on the aspect ratio and initial orientation of the inclusion as well as the components of the macroscopic deformation gradient. It should be noted that the rotation does not depend on $s$ for Neo-Hookean materials in the near-incompressible regime.
The rotation law is mechanical: it gives the rigid-body rotation of the inclusion selected by the surrounding elastic matrix. Once this rotation is known, the hard-magnetic remanent magnetization is treated as a material vector attached to the particle frame. Thus the rotation of the magnetic inclusion from initial magnetization $\bfM_0$ to final $\bfm$, accounting for the initial inclusion orientation and subsequent loading-induced rotation, is of the form $\bfm = \bfR^{\text{MIA}}\bfm_0$. The magnetic easy/remanent direction need not coincide with the geometric major axis; choosing them to coincide in the examples below is only an illustrative specialization and does not affect the mechanical rotation law. The magnetization considered here is residual remanent magnetization in high-coercivity particles, not induced magnetization in a soft-magnetic filler. Its magnitude is therefore taken to remain constant for the mechanical zero-field calculations.
Explicitly, if the initial remanent direction is $\beta$, then
\begin{equation}
    \bfm_0=m_0
    \begin{bmatrix}\cos\beta\\ \sin\beta\end{bmatrix},
    \qquad
    \bfm=m_0
    \begin{bmatrix}
        \cos(\beta+\theta^{\mathrm{MIA}})\\
        \sin(\beta+\theta^{\mathrm{MIA}})
    \end{bmatrix}.
    \label{eq:distinct-geometric-magnetic-angles}
\end{equation}
The mechanical rotation $\theta^{\mathrm{MIA}}$ depends on $\alpha_{\mathrm g}$ through \eqref{eq:rotation_map}, but the initial magnetic direction $\beta$ is an independent datum. Only the aligned examples specialize to $\beta=\alpha_{\mathrm g}=\alpha$.

\begin{figure}[htb!]
    \centering
    \includegraphics[width=0.49\textwidth]{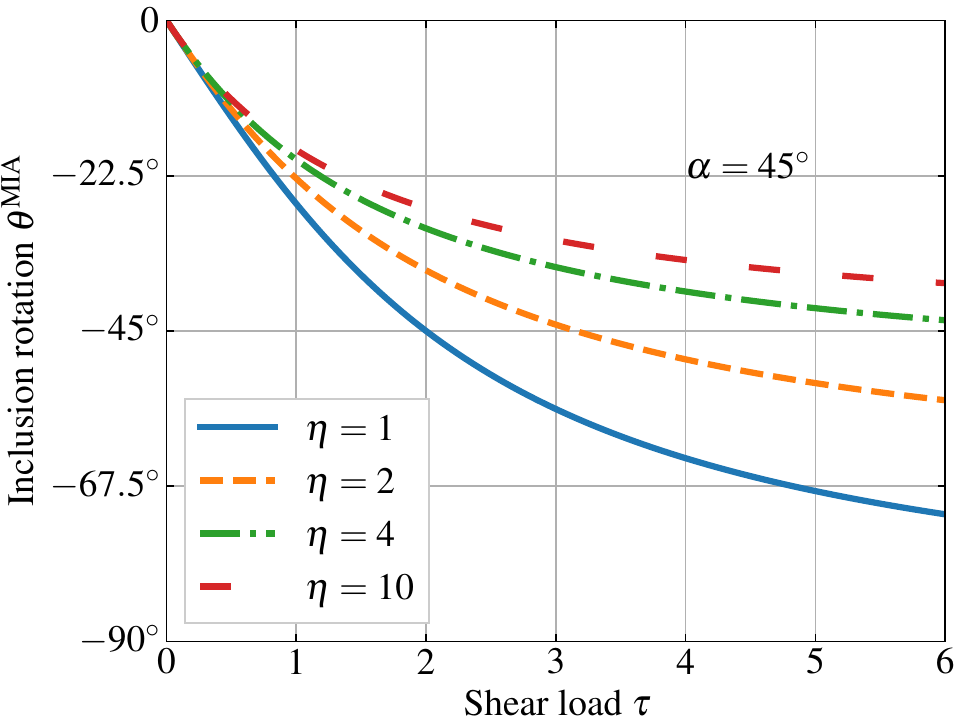}
    \caption{Rotation of the single inclusion in simple shear loading as the aspect ratio of the elliptical inclusion is changed from $\eta = 1$ to $\eta = 10$ starting with an initial orientation of the major axis at $\alpha = 45^{\circ}$. As the aspect ratio increases, all curves converge to a curve with maximum rotation of $45^{\circ}$.}
    \label{fig:shear_ar}
\end{figure}

\begin{figure}[htb!]
    \centering
    \includegraphics[width=0.49\textwidth]{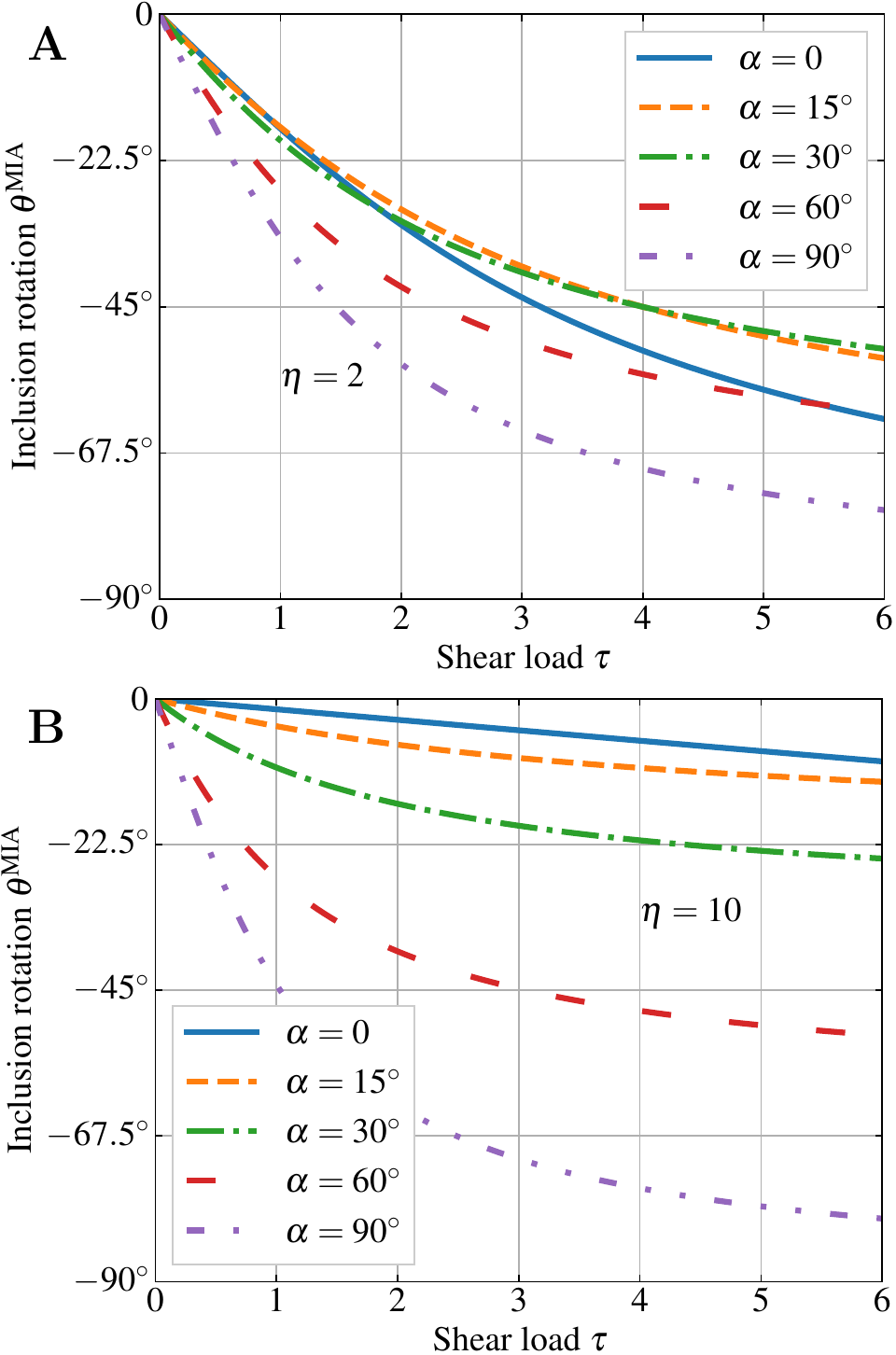}
    \caption{
        Rotation of a single inclusion in simple shear as the initial orientation of its major axis is varied from $\alpha=0^{\circ}$ to $\alpha=90^{\circ}$ for (A) $\eta=2$ and (B) $\eta=10$. At the lower aspect ratio there is no clear monotonic pattern with orientation, whereas at the higher aspect ratio the rotation increases more systematically with initial orientation.
    }

    \setcounter{subfigure}{0}
    \refstepcounter{subfigure}
    \label{fig:shear_comp_angle_ar2}
    \refstepcounter{subfigure}
    \label{fig:shear_comp_angle_ar10}
\end{figure}

The ``MIA rotation" depends on the inclusion aspect ratio and initial orientation. In \figref{fig:shear_ar}, the change of the predicted response with change in aspect ratio of the inclusion from $\eta = 1$ to $\eta = 10$ is shown. There is a decrease in the rotation that the inclusion undergo. Eventually, all curves converge following a clear pattern, and the maximum orientation achieved after convergence is $45^{\circ}$. In \figref{fig:shear_comp_angle_ar2} and \figref{fig:shear_comp_angle_ar10}, the orientation of the major axis of the inclusion is varied from $\alpha= 0^{\circ}$ to $\alpha = 90^{\circ}$ for a fixed aspect ratio of $\eta = 2$ and $\eta = 10$ , respectively. We observe a change the rotation for each case, with the maximum rotation occurring at $\alpha = 90^{\circ}$. For the case of $\eta = 2$, there is no clear pattern of increase in the rotation. Initially, when $\alpha = 0^{\circ}$, the inclusion rotates less, then as the load is increased, the material rotates at a faster rates than higher angles. However, a clear pattern of increasing rotation forms as we increase the initial orientation, as shown in \figref{fig:shear_comp_angle_ar10}.

Several models have been proposed to describe the rotation of rigid magnetic inclusion for magneto-rheological elastomers at macroscopic scale.\cite{GALIPEAU20131065} explored magneto-rheological elastomer composites with fibers of cylindrical cross-section. More recently, \cite{RAHMATI2023105136, Zhao2019MechanicsHardMagneticSoft} modeled magnetization in the reference configuration using the pull-back
\begin{align}
    \bfM_0 = J \barF^{-1} \bfm
    \label{eq:affine_mapping}
\end{align}
in which the residual magnetization is expected to change direction and magnitude with respect to deformation while accounting for change in volume. For the incompressible setting considered here, a magnitude-preserving comparison can be written directly in terms of the independently prescribed initial remanent vector $\bfm_0$, whose direction is $\beta$, as
\begin{align}
    \bfm = \frac{\barF \bfm_0}{J |\barF \bfm_0|}.
    \label{eq:rotation_Affine}
\end{align}
which will allow for a fair comparison to the model we developed, avoid the stretching of $\bfM_0$ that would generally be present in \eqref{eq:affine_mapping}. We will denote $\bfR^{\text{MM}}$ the rotation that the inclusion undergoes under the material mapping in \eqref{eq:rotation_Affine}.  Other researchers have argued that the magnetization ought to rotate under deformation owing to the rigidity of magnetic particles \cite{Stewart2023Magneto, danas2024stretch, Ca2023soft, Wang2020HardMagneticElastica}. This consideration differentiates this model from the material mapping model in \eqref{eq:affine_mapping} as it addresses the problem of stretching the magnetic field with deformation. It was then postulated to map the residual magnetization with rotation from polar decomposition \cite{danas2024stretch, YAN2023105095, Stewart2023Magneto,Wang2020HardMagneticElastica}.In the absence of an external field, the mapping is of the form
\begin{align}
    \bfm = \bfR^{\text{PD}}\bfm_0
    \label{eq:rotation_polarD}
\end{align}
where $\bfR^{\text{PD}}$ is the rotation from polar decomposition of $\barF$ such that $\barF = \bfR^{\text{PD}} \bfU$, and $\bfU$ is the stretch tensor. In this formulation is presented with a magnetic susceptibility being a scalar, hence the magnetization in the deformed configuration is the transformation of the auxiliary field $\mathcal{H}^r$, defined in \cite{danas2024stretch} as a remnant internal variable, and the rotation from polar decomposition.   

\subsection{Limit Cases}
The dependence of the result in \eqref{eq:rotation_map} on the inclusion aspect ratio can be eliminated by taking a limit $\eta \rightarrow 1$ for circular inclusion and $\eta \rightarrow \infty$ which can correspond, in some sense, to long and thin rigid inclusion. 
In the case of circular inclusion limit, the rotation in \eqref{eq:rotation_map} simply reduces to:
\begin{align}
\theta^{\text{MIA}}
  &= \tan^{-1}\left[
     \frac{%
       \begin{aligned}F_{12}-F_{21}
       \end{aligned}
       }{%
     \begin{aligned}F_{11}+F_{22}
       \end{aligned}
     }{
     }
     \right]
     \label{eq:rotation_map_circle}
\end{align}
in which the initial orientation of the inclusion or the magnetization does not affect the rotation, and the limit $\eta \rightarrow \infty$ becomes:
\begin{align}
\theta^{\text{MIA}}
  &= \tan^{-1}\left[
     \frac{%
       \begin{aligned}F_{11} \cot (\alpha ) +F_{12} -\cot (\alpha ) \left(F_{21} \cot (\alpha )+F_{22}\right)
       \end{aligned}
       }{%
     \begin{aligned}\cot (\alpha ) \left(F_{11} \cot (\alpha )+F_{12}+F_{21}\right)+F_{22}
       \end{aligned}
     }{
     }
     \right].
     \label{eq:rotation_map_lam}
\end{align}

It can be shown, by some straightforward algebra, that the limit case in \eqref{eq:rotation_map_lam} recovers the material mapping in \eqref{eq:rotation_Affine}, i.e. $\bfR^{\text{MM}} = \bfR^{\text{MIA}}$ when $\eta \rightarrow \infty$. From this perspective, the material mapping is a limit (special) case of our model, when the rigid inclusion is assumed long and thin.

\subsubsection{Simple shear and uniaxial tension in plane strain problem}

For this analysis, we consider two-dimensional rectangular body under plane strain loading. The boundary conditions are assumed to be affine on the boundary of the form $\bfy = \barF \bfx$. Two types of tests will be analyzed, namely, simple shear and tension tests. The macroscopic deformation gradients
\begin{align}
   \barF = \begin{pmatrix}
        \la & \tau \\
        0 & \la^{-1}
    \end{pmatrix}
    \label{eq:loading}
\end{align}
where $\la$ is the stretch in the $x-$direction while $\tau$ is shear. We are interested in comparing the rotation of the magnetic inclusion to the methods discussed above in \eqref{eq:rotation_Affine} and \eqref{eq:rotation_polarD} as these have been used by many researchers. We will also compare to results from finite element model.

\section{Comparison with Numerical Results} \label{sec:response}
In this section, we compute the mechanical response of the rigid inclusion subjected to mechanical loading in shear and tension. The response will be compared to numerical simulation as well as other proposed models for mapping the magnetization of rigid inclusions. It is worth noting that no external magnetic field is applied, rather, the focus is on evaluating large deformation models with magnetic inclusions. For the plots below we set the remanent magnetization to be aligned with the major axis of the elliptic inclusion for simplicity. This alignment is not an assumption of the theory: the inclusion orientation $\alpha$ determines the mechanical rotation, while any particle-fixed remanent magnetization direction is then rotated by the same rigid-body motion. In the distinct-angle notation introduced above, this plotting specialization is $\beta=\alpha_{\mathrm g}=\alpha$.

We highlight that the finite-element calculations are used as full-field numerical validation of the single-inclusion mechanical problem. 
They are not used to fit the MIA formula; the closed-form expression follows independently from the variational upper-bound calculation.

\subsection{Numerical modeling}

The rigid inclusion is modeled as an near-incompressible hyperelastic material, with a Neo-Hookean stored energy in an open-source finite element package FEniCSx \cite{alnaes2015fenics,logg2010dolfin}.The inclusion was also modeled as hyperelastic material for simplicity with constraints 
\begin{equation}
    \frac{\kappa}{2} ( \det \bfF  - 1 )^2 \quad \text{and} \quad \frac{\gamma}{2} | \bfF^T \bfF - \bfI |^2
\end{equation}
for near-incompressibility and for enforcing that the deformation gradient in the inclusion is a rotation respectively. 
The penalty constants were set to $\kappa = 10^6\mu$ and $\gamma = 10^7\mu$, with the energy normalized by the matrix shear modulus $\mu$. The deformation was modeled using first-order Lagrange elements in both phases. Additional finite-element details, including the quadrature rule, nonlinear solver tolerances, and extraction of the inclusion rotation, are provided in the Supplemental Material.

A sufficiently large domain was modeled to mimic the response of a rigid inclusion undergoing the effect of of macroscopic deformation. Both circular and elliptic inclusions were considered, while varying the angle of initial orientation of the inclusion. The affine boundary conditions were applied uniformly on the entire boundary. 
\begin{figure}[htb!]
\centering
    \includegraphics[width=0.9\columnwidth]{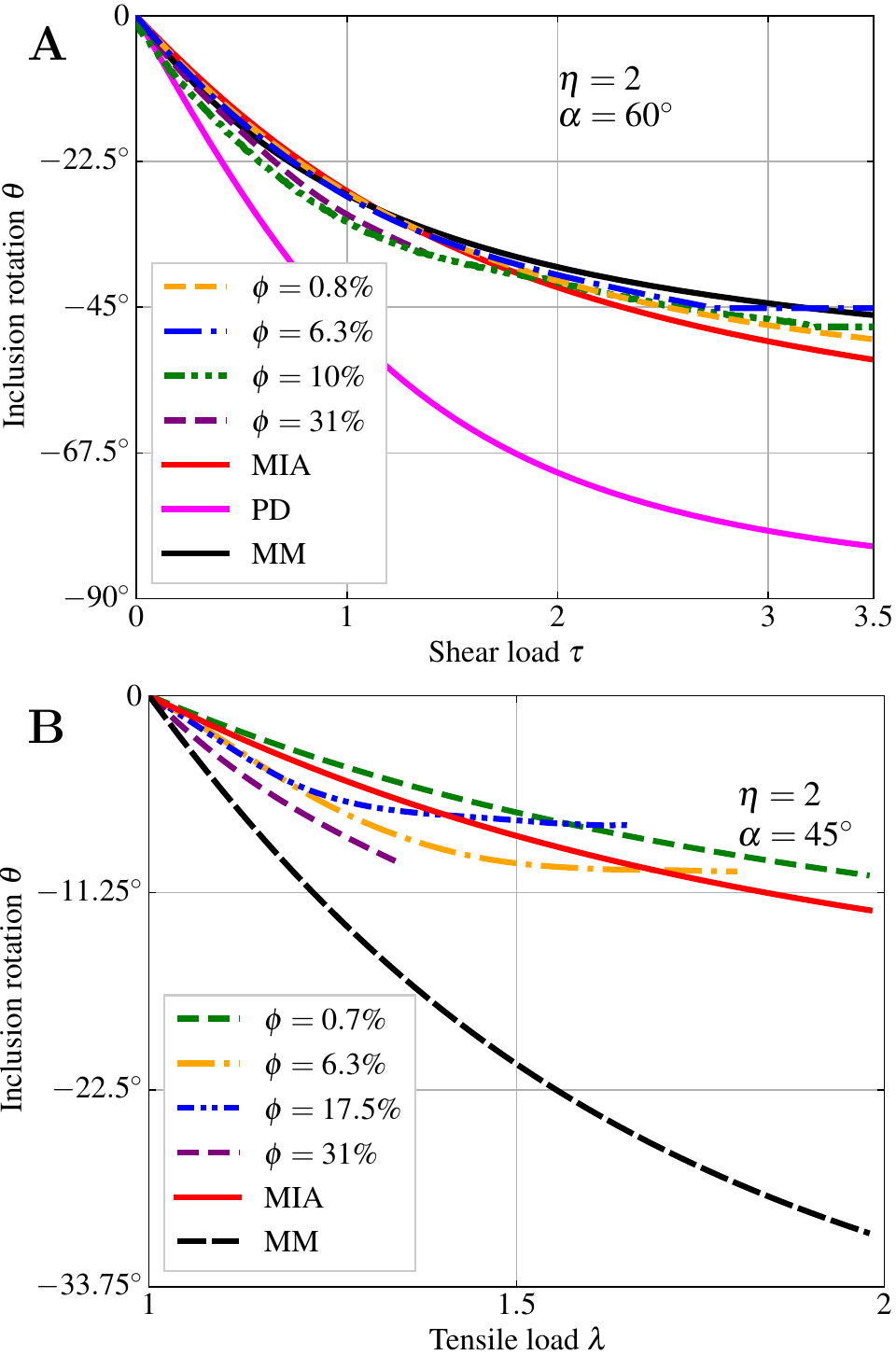}
    
    \caption{Single-inclusion finite-domain sensitivity for (A) shear and (B) uniaxial tension. Each panel compares the MIA rotation with the finite-element result as the inclusion volume fraction $\phi$ is changed by varying the size of the one-inclusion computational cell.}
    \label{fig:si_volume_fraction_sensitivity}
    \setcounter{subfigure}{0}
    \refstepcounter{subfigure}
    \label{fig:volume_fraction_sensitivity_shear}
    \refstepcounter{subfigure}
    \label{fig:volume_fraction_sensitivity_tensile}
\end{figure}

The finite-domain and volume-fraction sensitivity study is reported \figref{fig:si_volume_fraction_sensitivity}. Qualitatively, the shear curves remain comparatively close to the MIA over the tested range, whereas the tensile curves at the higher tested fractions ($10\%$ and $31\%$) departs relatively slightly from MIA compared to material mapping. Varying the size of a one-inclusion cell probes finite-domain and volume-fraction sensitivity; it does not model particle--particle mechanical interactions or magnetostatic interactions.

\subsection{Simple Shear}
In the case of simple shear loading, the tensile load $\lambda$ in \eqref{eq:loading} set to 1. The components of the "MIA rotation" in the form of $\cos$ and $\sin$ are written as

\begin{align}
\cos \theta^{\text{MIA}} &= \frac{\left(\eta ^4-1\right) \tau  \sin (2 \alpha )+2 \left(\eta ^4+6 \eta ^2+1\right)}{\sqrt{%
\begin{aligned}
    &\tau ^2 \left(-\left(\eta ^4-1\right) \cos (2 \alpha )+\eta ^4+6 \eta ^2+1\right)^2 \\
    &+\left(\left(\eta ^4-1\right) \tau  \sin (2 \alpha )+2 \left(\eta ^4+6 \eta ^2+1\right)\right)^2
\end{aligned}
}} \label{eq:shear_cos}\end{align}
\begin{align}
\sin \theta^{\text{MIA}} &= \frac{\tau  \left(-\left(\eta ^4-1\right) \cos (2 \alpha )+\eta ^4+6 \eta ^2+1\right)}{\sqrt{%
\begin{aligned}
    &\tau ^2 \left(-\left(\eta^4-1\right) \cos (2 \alpha )+\eta^4+6 \eta^2+1\right)^2 \\
    &+\left(\left(\eta^4-1\right) \tau  \sin (2 \alpha )+2 \left(\eta^4+6 \eta^2+1\right)\right)^2
\end{aligned}
}}.\label{eq:shear_sin}
\end{align}

\begin{figure}[htb!]
    \centering
    \includegraphics[width=0.9\columnwidth]{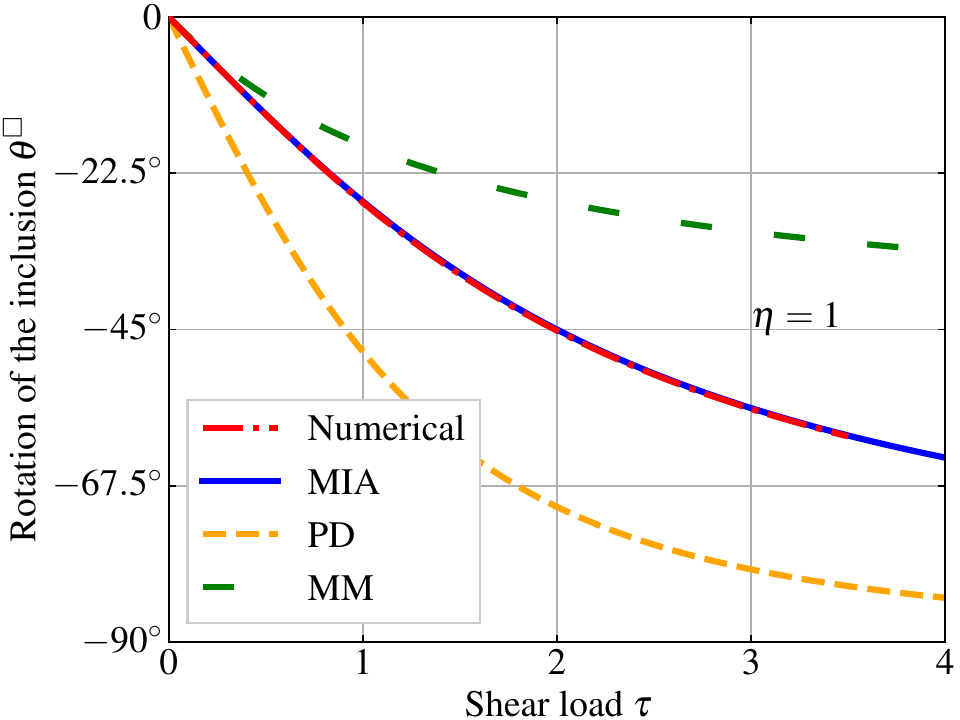}
    \caption{ Comparison of the ``MIA'' rotation to material mapping and polar decomposition rotations in \eqref{eq:rotation_Affine} and \eqref{eq:rotation_polarD}, respectively, under simple shear $\tau $ for a circular inclusion. The ``MIA'' rotation gives a better prediction.}
    \label{fig:shear_comp_circle}
\end{figure}

\figref{fig:shear_comp_circle} shows the results for a single circular rigid inclusion. The ``MIA'' model in \eqref{eq:rotation_map_circle} gives the best approximation of the mechanical response predicted by the numerical simulation. The polar decomposition over-predicts the response while the material mapping under-predicts. The polar decomposition method will be more accurate for softer inclusions, where the deformation map will tend to be homogeneous at the microscale. 

\begin{figure}[htb!]
    \centering
    \includegraphics[width=0.9\columnwidth]{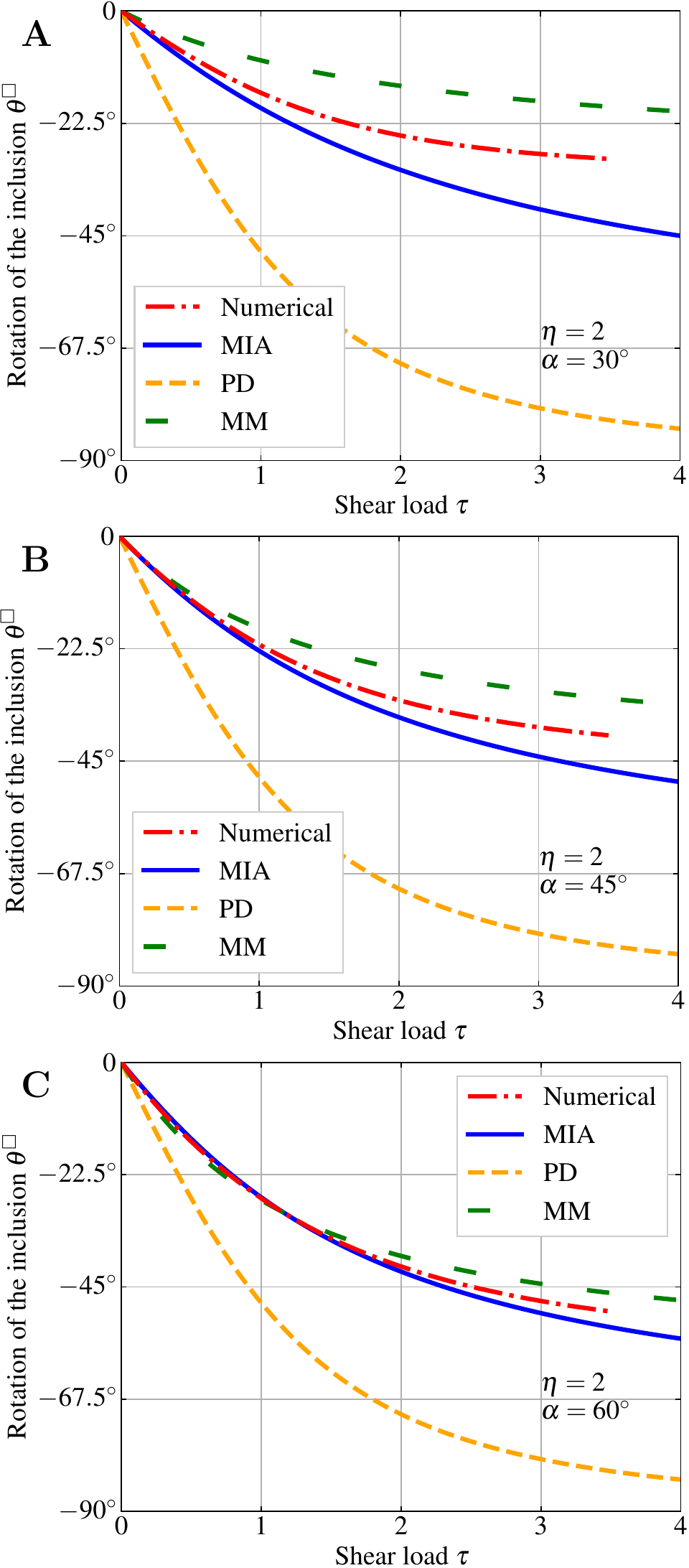}

    \label{fig:elliptic_shear}
    \caption{
        Comparison of the ``MIA'' rotation to material mapping and polar decomposition rotations in equations \eqref{eq:rotation_Affine} and \eqref{eq:rotation_polarD}, respectively, under simple shear $\tau$. The aspect ratio is fixed at $\eta = 2$. The initial orientation of the major axis of the elliptic inclusion is orientated at (A) $\alpha = 30^{\circ}$, (B) $\alpha = 45^{\circ}$ and (C) $\alpha = 60^{\circ}$.
    }

    \setcounter{subfigure}{0}
    \refstepcounter{subfigure}
    \label{fig:shear_comp30}
    \refstepcounter{subfigure}
    \label{fig:shear_comp45}
    \refstepcounter{subfigure}
    \label{fig:shear_comp60}
\end{figure}

\figref{fig:shear_comp30}, \figref{fig:shear_comp45} and \figref{fig:shear_comp60} presents the results for an elliptic inclusion for a fixed aspect ratio of $\eta = 2$ at various angles. In \figref{fig:shear_comp30}, the initial orientation of the inclusion is set to $30^\circ$, then the shear load is increased up to $\tau = 4$. The ``MIA'' rotation gives the prediction with the least error, while the affine rotation under-predicts and polar decomposition over-predicts. These results are similar to what was observed in the case of circular inclusion, except for the error for ``MIA'' increasing when the load is increased. \figref{fig:shear_comp45} and \figref{fig:shear_comp60} give similar results with improved predictions for all three methods. The ``MIA'' model still gives the best prediction. 

In the limiting cases of circular inclusion ($\eta \rightarrow 1$) and laminates ($\eta \rightarrow \infty$), the results in \eqref{eq:shear_cos} and \eqref{eq:shear_sin} become
\begin{align}
\cos \theta^{\text{MIA}} = \frac{2}{\sqrt{ 4 + \tau^2}}, 
\quad  \sin \theta^{\text{MIA}} = \frac{\tau}{\sqrt{ 4 + \tau^2}},
\end{align} 
\begin{align}
   \cos \theta^{\text{MIA}} = \frac{\tau  \sin (\alpha ) \cos (\alpha ) +1}{\sqrt{
    \tau ^2 \sin ^2(\alpha )+\tau  \sin (2 \alpha ) +1
}}, 
\end{align}
\begin{align}
\sin \theta^{\text{MIA}} = \frac{\tau  \sin ^2(\alpha )}{\sqrt{
    \tau ^2 \sin ^2(\alpha )+\tau  \sin (2 \alpha )+1}}
\end{align}
respectively. This recovers the linearized result 
\cite{SIBONI2012205}.

\subsection{Tension test}
The rotation of the rigid rotation as predicted by the ``MIA'' approach under a tension test is of the form
\begin{align}
\label{eq:tension-cos}
\cos \theta^{\text{MIA}} &= \frac{%
\begin{aligned}
\left(\eta ^4-1\right) \left(\lambda -\la^{-1}\right)& \cos (2 \alpha ) \\
&+\left(\eta ^4+6 \eta ^2+1\right) \left(\lambda +\la^{-1}\right)    
\end{aligned}}{\sqrt{%
\begin{aligned}
    &\left( \left(\eta^4-1\right) \left(\lambda - \la^{-1}\right) \sin (2 \alpha ) \right)^2\\
    &+\left( \left(\eta^4-1\right) \left(\lambda -\la^{-1}\right) \cos (2 \alpha ) \right. \\
    &+ \left.\left(\eta^4+6 \eta^2+1\right) \left(\lambda +\la^{-1}\right) \right)^2
\end{aligned}
}}%
\end{align}
\begin{align}
\label{eq:tension-sin}
\sin \theta^{\text{MIA}} &= \frac{\left(\eta ^4-1\right) \left(\lambda -\la^{-1}\right) \sin (2 \alpha )}{\sqrt{%
\begin{aligned}
    &\left( \left(\eta^4-1\right) \left(\lambda - \la^{-1}\right) \sin (2 \alpha ) \right)^2\\
    &+\left(\left(\eta^4-1\right) \left(\lambda -\la^{-1}\right) \cos (2 \alpha ) \right. \\
    &+ \left.\left(\eta^4+6 \eta^2+1\right) \left(\lambda +\la^{-1}\right)\right)^2
\end{aligned}
}}
\end{align}
where we have set the shear load $\tau = 0$ in \eqref{eq:loading}. In the results presented below, we only considered tension tests in the $x-$direction. The tests were run up to a tensile load of $\lambda = 2$. Also, the aspect ratio of the inclusions was fixed to $\eta = 2$ for all tests presented.  

\begin{figure}[htb!]
    \centering
    \includegraphics[width=0.9\columnwidth]{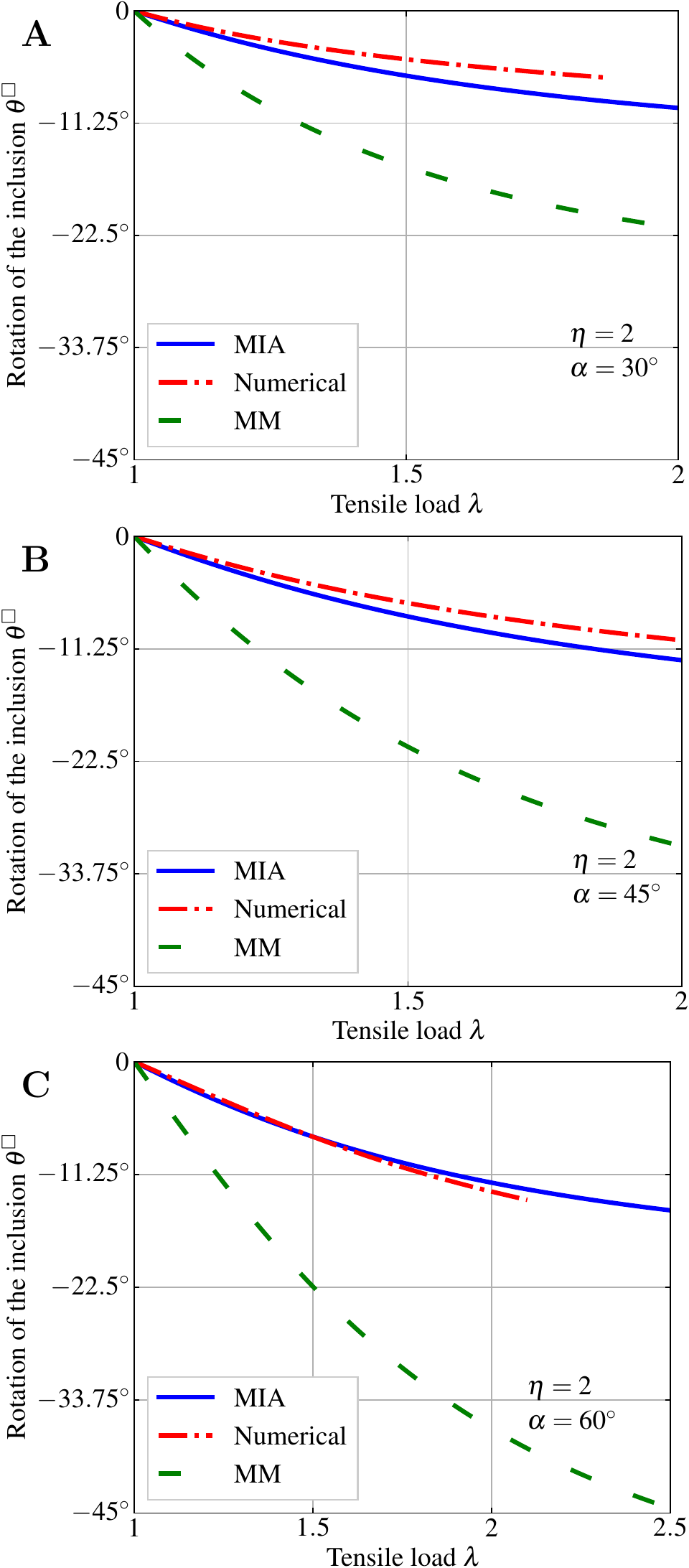}
    \caption{
        Comparison of the ``MIA'' rotation to material mapping and polar decomposition rotations in equations \eqref{eq:rotation_Affine} and \eqref{eq:rotation_polarD}, respectively, under  tension $\lambda$. The aspect ratio is fixed at $\eta = 2$. The initial orientation of the major axis of the elliptic inclusion is orientated at (A) $\alpha = 30^{\circ}$, (B) $\alpha = 45^{\circ}$ and (C) $\alpha = 60^{\circ}$.
    }

    \setcounter{subfigure}{0}
    \refstepcounter{subfigure}
    \label{fig:tension_comp30}
    \refstepcounter{subfigure}
    \label{fig:tension_comp45}
    \refstepcounter{subfigure}
    \label{fig:tension_comp60}
\end{figure}
At the dilute volume fraction used for \figref{fig:tension_comp30}, \figref{fig:tension_comp45}, and \figref{fig:tension_comp60}, the MIA rotation agrees well with the numerical results relative to the material mapping. However, the agreement at higher volume fractions is not quite as good (\figref{fig:volume_fraction_sensitivity_tensile}). The size of the interpolation region, $s$, was not computed explicitly here, but it can be obtained in a similar manner as the rotation; through energy minimization. 

The polar decomposition predicts zero rotation for all the elliptic inclusion tests presented here, since the deformation gradient is defined along the principal directions. This highlights a major limitation of the polar decomposition–based model in \cite{danas2024stretch} for inclusions under tension or compression. While the model provides accurate predictions for circular inclusions, it fails when the inclusion has an aspect ratio greater than one. The zero rotation reported in \cite{danas2024stretch} arises from the use of circular inclusions, and such results do not hold for elliptic inclusions, as demonstrated in our analysis.

\section{Emergent Magneto-Mechanical Coupling: Resolving Prior Model Pathologies}

Finally, to illustrate that our approach resolves inconsistencies and unphysical predictions of prior models, we consider a simple example of a magnetic field applied to a hard magnetic soft material. 
We focus on the device scale, where it is infeasible to resolve individual particles, and contrast the predictions of the first-principles model derived in this paper with those of existing ones in the literature.

The applied field should drive a mechanical deformation -- the very \textit{raison d'etre} for hard magnetic soft materials! -- but we see that the polar decomposition mapping predicts that no such coupling occurs in a simple example.
For transparency and to focus on the key physics, we make several simplifications below. 

Consider a specimen whose initial remanent magnetization is oriented at angle $\beta$ to a Cartesian basis, so that $\bfm_0=m_0[\cos\beta,\sin\beta]^T$. The particle geometry has the independent initial orientation $\alpha_{\mathrm g}$. In the aligned analytical example below, we specialize to $\beta=\alpha_{\mathrm g}=\alpha$. The numerical field-response figure is a separate aligned configuration with $\beta=\alpha_{\mathrm g}=0$, $\eta=2$, and $\bfH^{\mathrm{app}}$ directed along $-\bfe_2$.
The externally applied field $\bfH^{\mathrm{app}}$ is directed along $\bfe_1$. We use the hard-magnetic idealization throughout: the elastomer is mechanically soft, while the particles have high coercivity and a remanent magnetization that remains locked to the particle frame for fields below the coercive or reprogramming scale. Thus the applied field does not introduce an internal magnetization-rotation variable inside the particle; instead, it exerts a torque on the magnetized rigid inclusion, and the soft matrix deforms in response. For simplicity, we assume that there is no induced magnetization beyond the residual magnetization.

\begin{figure}[htb!]
    \centering
    \includegraphics[width=0.95\columnwidth]{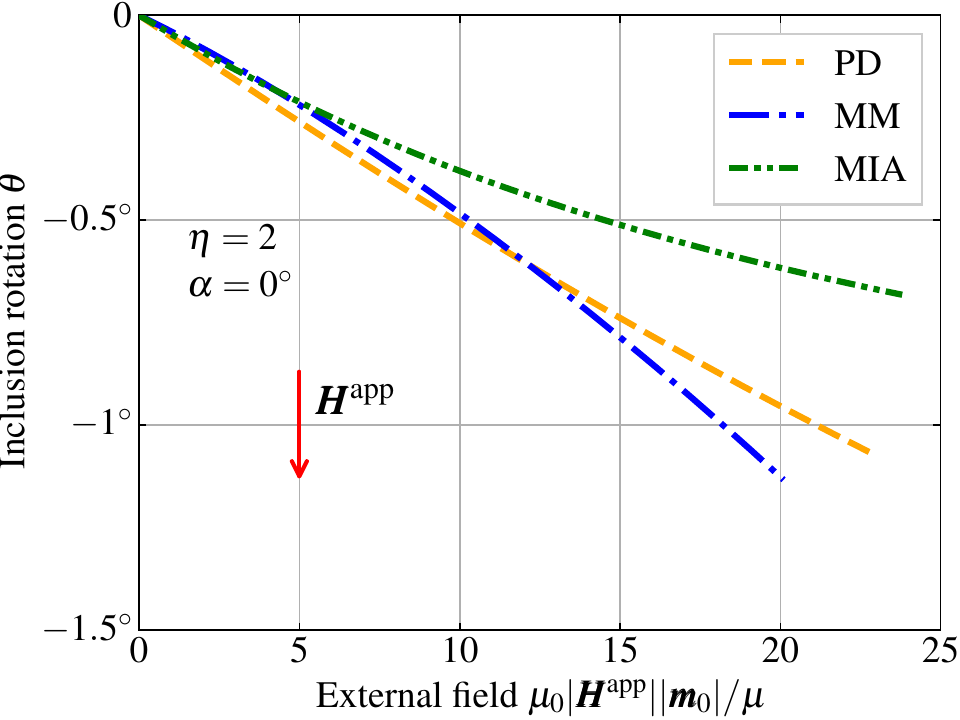}
    \caption{Rotation versus applied field predicted by the polar-decomposition (PD), material-mapping (MM), and micromechanical-interpolation (MIA) models for $\beta=\alpha_{\mathrm g}=0$, inclusion aspect ratio $\eta=2$, and $\bfH^{\mathrm{app}}$ directed along $-\bfe_2$. Because the remanent magnetization is aligned with the particle axis in this example, each curve is also the change in magnetization direction; the remanent-magnetization magnitude is held fixed.}
    \label{fig:field-rotation}
\end{figure}

In thermodynamic terms, we minimize a free energy with Zeeman coupling to the externally applied field,
\begin{equation}
    \int \left(W(\nabla\bfy) - \mu_0 \bfH^{\mathrm{app}}\cdot\bfm\right).
\end{equation}
We neglect the magnetic field energy in this illustrative calculation because the torque-producing Zeeman term is the contribution relevant to the comparison among deformation-to-magnetization mappings. We also do not add a separate self-demagnetizing torque for an isolated uniformly magnetized ellipsoidal hard particle: the demagnetization tensor and the remanent magnetization rotate together in the particle frame, so this self-demagnetizing contribution is constant under rigid-body rotation of the particle. Demagnetizing fields can affect the stability and magnitude of the remanent magnetization, and magnetostatic dipole--dipole interactions can modify the local field in concentrated composites, but those effects are outside the single-inclusion mapping derived here and are absorbed into the hard-magnetic material idealization.

Assuming that the deformation and the magnetization are uniform implies that the integrand is constant, leading to a simple algebraic problem:
\begin{equation}
\label{eqn:magnetic-free-energy}
    \min_\bfF \left(W(\bfF) - \mu_0 \bfH^{\mathrm{app}}\cdot\bfm(\bfm_0,\bfF) \right)
\end{equation}
where $\bfm(\bfF,\bfm_0)$ is the mapping of $\bfm_0$ due to $\bfF$.

If we simply minimize \eqref{eqn:magnetic-free-energy} over all possible deformations, all of the models provide a simple and physical result: $\bfm$ aligns with $\bfH^{\mathrm{app}}$, and the deformation is a rotation.
We therefore constrain the deformation to obtain a non-trivial result; we use $\bfF = \begin{bmatrix}
    \lambda & 0 \\ 0 & \lambda^{-1}
\end{bmatrix}$
which is simple enough that the polar decomposition can be readily applied.
When restricted to a simple elongational deformation, the polar decomposition model \cite{danas2024stretch} yields a vacuous result: the rotation reduces to the identity matrix. This leads to a total decoupling of the magnetic Zeeman energy from the elastic energy, resulting in a prediction of zero induced deformation $\lambda = 1$  regardless of the field strength $\bfH^{\mathrm{app}}$ or particle orientation  $\alpha$. For a material whose entire \emph{raison d'être}  is to convert magnetic fields into motion, this null prediction represents a fundamental breakdown in physical fidelity.

However, the first-principles approach generically predicts coupling for an inclined anisotropic inclusion, while recovering zero coupling in the relevant symmetry cases, including $\eta=1$ or $\sin^2\alpha\cos\alpha=0$.
We use the MIA mapping \eqref{eq:rotation_map} in the energy \eqref{eqn:magnetic-free-energy} to obtain an algebraic function in terms of $\lambda$ which clearly shows the coupling between the applied field and the induced deformation.
While numerical minimization is straightforward, it requires specifying the values of the material properties; to obtain a generic solution, we linearize in the small-strain limit $\lambda \simeq 1 + \epsilon$ to obtain the induced strain:
\begin{equation}
\begin{aligned}
    \epsilon
    =
    -\frac{\mu_0 |\bfH^{\mathrm{app}}| |\bfm_0|}{2\mu}
    \frac{\eta^4-1}{\eta^4 + 6\eta^2 +1} \sin^2 \alpha \cos \alpha.
\end{aligned}
\end{equation}
where we have further simplified by assuming that $\frac{\mu_0 |\bfH^{\mathrm{app}}| |\bfm_0|}{2\mu} \ll 1$; the Supplemental Material provides the complete expression and details about the derivation.
The expression above shows that the first-principles approach captures the central aspects of magneto-mechanical coupling; we also see that the limit case $\eta=1$ recovers the polar decomposition model and wrongly predicts the lack of magneto-mechanical coupling. This distinction is not merely theoretical; it has profound implications for the rational design of soft actuators. At the device scale, a model that fails to capture geometry-induced rotation would predict a ``phantom'' device; one that remains static despite intense magnetic stimulation. 

A nonuniform applied field would add force terms associated with spatial gradients of $\bfH^{\mathrm{app}}$, which are important for navigation or translation of a device. Those terms are distinct from the local deformation-to-remanent-magnetization mapping $\bfm(\bfF)$ derived here. 
The closed-form law and the finite-element comparisons presented here are single-inclusion calculations. Changing the size of a one-inclusion computational cell tests finite-domain and volume-fraction sensitivity, but does not resolve explicit particle--particle mechanical interactions. Neither those many-particle mechanical interactions nor long-range magnetostatic interactions are included in the present model.

\section{Concluding Remarks} \label{sec:conclusion}

We addressed the central question of how to map the remanent magnetization of a rigid inclusion through finite deformation in a way that preserves magnitude yet captures the geometry-induced rotations.
Starting from an energetic micromechanical framework, we constructed a variational upper bound and used this derive a closed-form rotation for an individual rigid inclusion.
This yields an explicit constitutive mapping grounded in micromechanics rather than imposed in an ad hoc way, as in prior approaches.

Within its dilute, single-inclusion scope, the predicted inclusion rotation agrees closely with the numerical results under shear and tension. The one-inclusion cell-size study shows comparatively close shear curves over the tested range, while the visible separation of the higher-fraction tensile curves delimits the range in which the closed-form approximation is quantitatively reliable. These cell calculations do not establish agreement for many-particle mechanical or magnetostatic interactions.
Our approach is analytically transparent and accommodates the inclusion shape and initial orientation. 
These features enable it to serve as a useful forward model for the microstructural design of dilute hard magnetic soft materials.
The closed-form expression presented here is the two-dimensional plane-strain specialization. The same variational principle extends to three-dimensional ellipsoidal inclusions by minimizing over $\mathrm{SO}(3)$, and future work can combine this mechanical mapping with explicit magnetostatic interactions or field-gradient forces in concentrated, device-scale settings.

Finally, our fundamental approach shows that prior \emph{ad hoc} approaches that were seemingly conflicting are, in fact, simply special limit cases of our general approach.

\section{Acknowledgments}
    Kaushik Dayal, Daniel Katusele, and Carmel Majidi acknowledge support from NSF (DMS 2108784, DMREF 1921857), BSF (2018183), and ARO (MURI W911NF-24-2-0184).
    Daniel Katusele acknowledges support from the CMU TCS Presidential Fellowship.
    Pradeep Sharma acknowledges support from the Cullen Professorship.
    
\section{Competing interest}
The authors declare that they have no competing interests.

\section{Funding}
This work is supported in part by funds from NSF (DMS 2108784, DMREF 1921857), BSF (2018183), and ARO (MURI W911NF-24-2-0184).

\section{Author contributions}


K.D. and P.S. conceived the study.
K.D. formulated the overall approach. 
D.K. and P.S. performed the theoretical calculations.
D.K. performed the numerical calculations and prepared the original draft of the manuscript, which was revised by K.D. and P.S.
K.D., P.S., and C.M. supervised the research efforts of D.K.
L.L. contributed to general technical discussions.

\section{Data availability}

A version of the code developed for this work is available at \cite{githubGitHubDanielkatuseleMIA}:
\url{https://github.com/danielkatusele/Micromechanical_Interpolation_Approximation.git}

\bibliography{micromechanics}
\bibliographystyle{oup-plain}

\clearpage



\section*{Supplementary material}

# Supplementary Information: How Does Hard Magnetic Soft Matter Deform?

**Daniel Katusele,[1,2] Carmel Majidi,[3] Liping Liu,[4] Pradeep Sharma[5] and Kaushik Dayal[2,3,6,*]**

[1]Program in Computational Mechanics, Carnegie Mellon University
[2]Department of Civil and Environmental Engineering, Carnegie Mellon University
[3]Department of Mechanical Engineering, Carnegie Mellon University
[4]Department of Mathematics, Rutgers University
[5]Department of Mechanical Engineering, University of Houston
[6]Center for Nonlinear Analysis, Carnegie Mellon University
*To whom correspondence should be addressed: Kaushik.Dayal@cmu.edu

# 1. Derivation of Ansatz for MIA Mapping

The ansatz behind the Micromechanical Interpolation Approach (MIA) mapping is derived by formulating a trial field over the micromechanical domain which satisfy boundary conditions. The micromechanical domain is split into three subdomains, the inclusions, the matrix around the inclusion and the matrix outside the inclusion.

## 1.1. Trial deformation in the three subdomains

The domain is partitioned by the level-set function $z(\boldsymbol{x})$ into three regions:

$$\Omega_i = \{\boldsymbol{x} : \ z(\boldsymbol{x}) \leq 1\}, \quad \Omega_p = \{\boldsymbol{x} : \ 1 < z(\boldsymbol{x}) < s^2\}, \quad \Omega_m = \{\boldsymbol{x} : \ z(\boldsymbol{x}) \geq s^2\}.$$

The parameter $s > 1$ defines the outer boundary of the interpolation region. If $z = 1$ is the inclusion boundary, then $z = s^2$ is a geometrically similar outer level set. For an elliptical inclusion this means that the outer semi-axes are $s$ times the inclusion semi-axes. The parameter is part of the admissible trial field used to construct the variational upper bound; it is not a fitted material constant. In the Neo-Hookean calculation leading to the closed-form planar rotation law, the stationarity condition for the rotation is independent of $s$, even though the numerical value of the upper-bound energy can depend on the size of the interpolation layer. The trial deformation is prescribed as a rigid deformation in $\Omega_i$ and as an affine deformation in $\Omega_m$:

$$\begin{aligned} \boldsymbol{y}(\boldsymbol{x}) &= \boldsymbol{R}_0\boldsymbol{x} + \boldsymbol{c}_0 \qquad &(\boldsymbol{x} \in \Omega_i), \\ \boldsymbol{y}(\boldsymbol{x}) &= \overline{\boldsymbol{F}}\boldsymbol{x} \qquad &(\boldsymbol{x} \in \Omega_m). \end{aligned}$$

In the interpolation region $\Omega_p$, we assume $\boldsymbol{y}$ depends linearly on $z$ so that it matches the two boundary values:

$$\begin{aligned} \boldsymbol{y}(\boldsymbol{x}) &= \boldsymbol{R}_0\boldsymbol{x} + \boldsymbol{c}_0 \qquad &\text{at} \quad z(\boldsymbol{x}) = 1, \\ \boldsymbol{y}(\boldsymbol{x}) &= \overline{\boldsymbol{F}}\boldsymbol{x} \qquad &\text{at} \quad z(\boldsymbol{x}) = s^2. \end{aligned}$$

A linear expression satisfying these constraints is

$$\boldsymbol{y}(\boldsymbol{x}) = \frac{(\boldsymbol{R}_0\boldsymbol{x} + \boldsymbol{c}_0)\,(z(\boldsymbol{x}) - s^2) - \overline{\boldsymbol{F}}\boldsymbol{x}\,(z(\boldsymbol{x}) - 1)}{1 - s^2}, \qquad \boldsymbol{x} \in \Omega_p,$$

Therefore, combining all three regions yields Eq. [4] in the manuscript:

$$\boldsymbol{y}(\boldsymbol{x}) = \begin{cases} \boldsymbol{R}_0\boldsymbol{x} + \boldsymbol{c}_0, & \boldsymbol{x} \in \Omega_i, \\ \dfrac{(\boldsymbol{R}_0\boldsymbol{x} + \boldsymbol{c}_0)\,(z(\boldsymbol{x}) - s^2) - \overline{\boldsymbol{F}}\boldsymbol{x}\,(z(\boldsymbol{x}) - 1)}{1 - s^2}, & \boldsymbol{x} \in \Omega_p, \\ \overline{\boldsymbol{F}}\boldsymbol{x}, & \boldsymbol{x} \in \Omega_m. \end{cases} \tag{1}$$

## 1.2. Computing the deformation gradient in the interpolation region

In $\Omega_p$ we start from

$$\boldsymbol{y}(\boldsymbol{x}) = \frac{(\boldsymbol{R}_0\boldsymbol{x} + \boldsymbol{c}_0)\,(z - s^2) - \overline{\boldsymbol{F}}\boldsymbol{x}\,(z-1)}{1 - s^2}, \qquad \boldsymbol{x} \in \Omega_p,$$

where, for brevity, $z = z(\boldsymbol{x})$. We proceed with the derivation in indicial notation. Write components $y_i$ and differentiate with respect to $x_j$:

$$\frac{\partial y_i}{\partial x_j} = \frac{1}{1-s^2}\left[\frac{\partial}{\partial x_j}\Big((\boldsymbol{R}_0\boldsymbol{x} + \boldsymbol{c}_0)_i\,(z - s^2)\Big) - \frac{\partial}{\partial x_j}\Big((\overline{\boldsymbol{F}}\boldsymbol{x})_i\,(z-1)\Big)\right].$$

Apply the product rule. First note:

$$\frac{\partial}{\partial x_j}(\boldsymbol{R}_0\boldsymbol{x} + \boldsymbol{c}_0)_i = (\boldsymbol{R}_0)_{ij}, \qquad \frac{\partial}{\partial x_j}(\overline{\boldsymbol{F}}\boldsymbol{x})_i = F_{ij}, \qquad \frac{\partial}{\partial x_j}(z) = \partial_j z.$$

Therefore,

$$\frac{\partial y_i}{\partial x_j} = \frac{1}{1-s^2}\left[(\boldsymbol{R}_0)_{ij}(z-s^2) + (\boldsymbol{R}_0\boldsymbol{x} + \boldsymbol{c}_0)_i\,\partial_j z - F_{ij}(z-1) - (\overline{\boldsymbol{F}}\boldsymbol{x})_i\,\partial_j z\right].$$

Group the $\partial_j z$ terms:

$$\frac{\partial y_i}{\partial x_j} = \frac{1}{1-s^2}\left[\Big((\boldsymbol{R}_0\boldsymbol{x} + \boldsymbol{c}_0)_i - (\overline{\boldsymbol{F}}\boldsymbol{x})_i\Big)\partial_j z + (\boldsymbol{R}_0)_{ij}(z-s^2) - F_{ij}(z-1)\right].$$

In tensor form, this gives Eq. [5]:

$$\widetilde{\boldsymbol{F}}(\boldsymbol{x}) = \nabla_{\boldsymbol{x}}\boldsymbol{y}(\boldsymbol{x}) = \frac{(\boldsymbol{R}_0\boldsymbol{x} + \boldsymbol{c}_0 - \overline{\boldsymbol{F}}\boldsymbol{x}) \otimes \nabla_{\boldsymbol{x}} z + \boldsymbol{R}_0(z-s^2) - \overline{\boldsymbol{F}}(z-1)}{1-s^2}, \qquad \boldsymbol{x} \in \Omega_p. \tag{2}$$

# 2. Rescaled polar coordinates for the interpolation-region boundaries

The level-set function and deformation gradient will be transformed to polar coordinates for easy computation. We will proceed with elliptic level-set functions centered at the origin and rescaled to a circle in polar coordinates.

## 2.1. Interpolation domain in rescaled polar coordinate

*Elliptic level-set.*
For a single ellipse with aspect ratio $\eta > 0$ and orientation $\alpha$, the level-set is

$$z(\boldsymbol{x}) = \boldsymbol{x}\cdot\boldsymbol{R}_\alpha^T\boldsymbol{P}_\eta\boldsymbol{R}_\alpha\boldsymbol{x}, \qquad \boldsymbol{P}_\eta = \begin{pmatrix}\frac{1}{\eta^2} & 0\\ 0 & 1\end{pmatrix}.$$

To perform the rescaling and polar-coordinate transformation cleanly, we work in the ellipse-aligned frame (i.e. after rotating coordinates so that $\alpha = 0$). Then $\boldsymbol{R}_\alpha = \boldsymbol{I}$ and

$$z(\boldsymbol{x}) = \boldsymbol{x}\cdot\boldsymbol{P}_\eta\boldsymbol{x} = \frac{x_1^2}{\eta^2} + x_2^2.$$

*Ellipse-to-circle rescaling.*
Define the rescaled coordinate $\boldsymbol{u}$ by

$$\boldsymbol{u} = \boldsymbol{P}_\eta^{1/2}\boldsymbol{x}, \qquad \boldsymbol{P}_\eta^{1/2} = \begin{pmatrix}\frac{1}{\eta} & 0\\ 0 & 1\end{pmatrix}.$$

Equivalently,

$$\boldsymbol{x} = \boldsymbol{P}_\eta^{-1/2}\boldsymbol{u}, \qquad \boldsymbol{P}_\eta^{-1/2} = \begin{pmatrix}\eta & 0\\ 0 & 1\end{pmatrix}.$$

*Transformation of the level-set.*
Starting from $z(\boldsymbol{x}) = \boldsymbol{x}^T\boldsymbol{P}_\eta\boldsymbol{x}$ and substituting $\boldsymbol{x} = \boldsymbol{P}_\eta^{-1/2}\boldsymbol{u}$, we obtain

$$\begin{aligned} z(\boldsymbol{x}) &= (\boldsymbol{P}_\eta^{-1/2}\boldsymbol{u})^T\,\boldsymbol{P}_\eta\,(\boldsymbol{P}_\eta^{-1/2}\boldsymbol{u})\\ &= \boldsymbol{u}^T\big(\boldsymbol{P}_\eta^{-1/2}\boldsymbol{P}_\eta\boldsymbol{P}_\eta^{-1/2}\big)\boldsymbol{u}\\ &= \boldsymbol{u}^T\boldsymbol{I}\,\boldsymbol{u} = |\boldsymbol{u}|^2. \end{aligned}$$

Hence

$$z(\boldsymbol{x}) = u_1^2 + u_2^2.$$

*Polar coordinates in the rescaled variable.*
Introduce polar coordinates for $\boldsymbol{u}$:

$$\boldsymbol{u} = r(\cos\phi, \sin\phi), \qquad u_1 = r\cos\phi, \quad u_2 = r\sin\phi, \qquad r \geq 0,\ \phi \in [0, 2\pi).$$

Then the original coordinate $\boldsymbol{x} = \boldsymbol{P}_\eta^{-1/2}\boldsymbol{u}$ becomes

$$x_1 = \eta\, u_1 = \eta\, r\cos\phi, \qquad x_2 = u_2 = r\sin\phi,$$

and the level-set becomes

$$z(\boldsymbol{x}) = |\boldsymbol{u}|^2 = r^2.$$

*Boundaries of the interpolation region in rescaled polar coordinates.*
By definition,

$$\Omega_p = \{x :\ 1 < z(x) < s^2\}.$$

Using $z(x) = r^2$, this becomes

$$\Omega_p = \{(r, \phi) :\ 1 < r^2 < s^2,\ \phi \in [0, 2\pi)\}.$$

Since $r \geq 0$, the inequalities $1 < r^2 < s^2$ are equivalent to

$$1 < r < s.$$

Thus, in rescaled polar coordinates, the interpolation region is the circular annulus

$$\Omega_p = \{(r, \phi) :\ 1 < r < s,\ \phi \in [0, 2\pi)\}.$$

## 2.2. Expression for the deformation gradient in rescaled polar coordinates

The translation is set to zero ($\boldsymbol{c}_0 = \boldsymbol{0}$) and the difference tensor is defined as

$$\Delta := \boldsymbol{R}_0 - \overline{\boldsymbol{F}}.$$

Then

$$\boldsymbol{R}_0\boldsymbol{x} + \boldsymbol{c}_0 - \overline{\boldsymbol{F}}\boldsymbol{x} = (\boldsymbol{R}_0 - \overline{\boldsymbol{F}})\boldsymbol{x} = \Delta\boldsymbol{x},$$

and the expression for $\widetilde{\boldsymbol{F}}$ becomes

$$\widetilde{\boldsymbol{F}}(\boldsymbol{x}) = \frac{(\Delta\boldsymbol{x}) \otimes \nabla_{\boldsymbol{x}} z(\boldsymbol{x}) + \boldsymbol{R}_0\big(z(\boldsymbol{x}) - s^2\big) - \overline{\boldsymbol{F}}\big(z(\boldsymbol{x}) - 1\big)}{1 - s^2}.$$

We rewrite the last two terms in the numerator in terms of $\overline{\boldsymbol{F}}$ and $\Delta$ without introducing any further rearrangements. Using $\boldsymbol{R}_0 = \overline{\boldsymbol{F}} + \Delta$,

$$\begin{aligned}
\boldsymbol{R}_0(z - s^2) - \overline{\boldsymbol{F}}(z - 1) &= (\overline{\boldsymbol{F}} + \Delta)(z - s^2) - \overline{\boldsymbol{F}}(z - 1) \\
&= \overline{\boldsymbol{F}}(z - s^2) + \Delta(z - s^2) - \overline{\boldsymbol{F}}(z - 1) \\
&= \big(\overline{\boldsymbol{F}}z - \overline{\boldsymbol{F}}s^2\big) + \Delta(z - s^2) - \big(\overline{\boldsymbol{F}}z - \overline{\boldsymbol{F}}\big) \\
&= \big(\overline{\boldsymbol{F}}z - \overline{\boldsymbol{F}}s^2 - \overline{\boldsymbol{F}}z + \overline{\boldsymbol{F}}\big) + \Delta(z - s^2) \\
&= \big(\overline{\boldsymbol{F}} - \overline{\boldsymbol{F}}s^2\big) + \Delta(z - s^2) \\
&= (1 - s^2)\overline{\boldsymbol{F}} + (z - s^2)\Delta.
\end{aligned}$$

Therefore,

$$\begin{aligned}
\widetilde{\boldsymbol{F}}(\boldsymbol{x}) &= \frac{(\Delta\boldsymbol{x}) \otimes \nabla_{\boldsymbol{x}} z(\boldsymbol{x}) + (z(\boldsymbol{x}) - s^2)\Delta + (1 - s^2)\overline{\boldsymbol{F}}}{1 - s^2} \\
&= \overline{\boldsymbol{F}} + \frac{(\Delta\boldsymbol{x}) \otimes \nabla_{\boldsymbol{x}} z(\boldsymbol{x}) + (z(\boldsymbol{x}) - s^2)\Delta}{1 - s^2}.
\end{aligned}$$

In the ellipse-aligned frame ($\alpha = 0$),

$$z(\boldsymbol{x}) = \frac{x_1^2}{\eta^2} + x_2^2.$$

Compute the gradient with respect to $x$:

$$\nabla_{\boldsymbol{x}} z(\boldsymbol{x}) = \begin{pmatrix} \dfrac{\partial}{\partial x_1}\left(\dfrac{x_1^2}{\eta^2} + x_2^2\right) \\ \dfrac{\partial}{\partial x_2}\left(\dfrac{x_1^2}{\eta^2} + x_2^2\right) \end{pmatrix} = \begin{pmatrix} \dfrac{2x_1}{\eta^2} \\ 2x_2 \end{pmatrix}.$$

Substitute $x_1 = \eta\, r\cos\phi$ and $x_2 = r\sin\phi$:

$$\nabla_{\boldsymbol{x}} z(\boldsymbol{x}) = \begin{pmatrix} \dfrac{2(\eta\, r\cos\phi)}{\eta^2} \\ 2(r\sin\phi) \end{pmatrix} = \begin{pmatrix} \dfrac{2\, r\cos\phi}{\eta} \\ 2\, r\sin\phi \end{pmatrix}.$$

Next compute $\varDelta\boldsymbol{x}$ using $\boldsymbol{x} = (\eta\, r\cos\phi,\ r\sin\phi)^T$:

$$\varDelta\boldsymbol{x} = \varDelta \begin{pmatrix} \eta\, r\cos\phi \\ r\sin\phi \end{pmatrix}.$$

Finally, since $z(\boldsymbol{x}) = r^2$, we have

$$z(\boldsymbol{x}) - s^2 = r^2 - s^2.$$

Start from the form

$$\widetilde{\boldsymbol{F}}(\boldsymbol{x}) = \overline{\boldsymbol{F}} + \frac{(\varDelta\boldsymbol{x}) \otimes \nabla_{\boldsymbol{x}} z(\boldsymbol{x}) + (z(\boldsymbol{x}) - s^2)\varDelta}{1 - s^2}.$$

Substitute the expressions for $\nabla_{\boldsymbol{x}} z(\boldsymbol{x})$, $\varDelta\boldsymbol{x}$ and $z(\boldsymbol{x}) - s^2$ derived above to yield the rescaled polar-coordinates form

$$\widetilde{\boldsymbol{F}}(r,\phi) = \overline{\boldsymbol{F}} + \frac{1}{1-s^2}\left(\varDelta \begin{pmatrix} \eta\, r\cos\phi \\ r\sin\phi \end{pmatrix} \otimes \begin{pmatrix} \dfrac{2\, r\cos\phi}{\eta} \\ 2r\sin\phi \end{pmatrix}\right) + \frac{(r^2 - s^2)\varDelta}{1 - s^2}$$
$$\text{for} \quad 1 < r < s,\ \phi \in [0, 2\pi).$$

# 3. Energy integral over the interpolation region

## 3.1. Energy functional

*Upper-bound energy.*

The manuscript introduces a Neo-Hookean upper-bound functional in the form

$$W(\widetilde{\boldsymbol{F}}) = \min_{\boldsymbol{R}_0, \boldsymbol{c}_0, s} \frac{1}{|\varOmega|}\left(\int_{\varOmega_p} \frac{\mu_m}{2}\operatorname{tr}\big(\widetilde{\boldsymbol{F}}^T\widetilde{\boldsymbol{F}} - \boldsymbol{I}\big) + \int_{\varOmega_m} \frac{\mu_m}{2}\operatorname{tr}\big(\overline{\boldsymbol{F}}^T\overline{\boldsymbol{F}} - \boldsymbol{I}\big)\, d\boldsymbol{x}\right).$$

Only the integral over $\varOmega_p$ depends on $\boldsymbol{R}_0$ (and hence on the rotation angle $\theta$).

*Reduced energy.*

Up to additive constants (and focusing on the $\theta$-dependent part), the derivation note reduces the problem to the minimization of

$$E(\theta) := \int_{\varOmega_p} \operatorname{tr}\big(\widetilde{\boldsymbol{F}}^T\widetilde{\boldsymbol{F}}\big)\, d\boldsymbol{x}.$$

In what follows we compute $E(\theta)$ explicitly as an integral over $\varOmega_p$.

## 3.2. Energy functional in rescaled polar coordinates

*Interpolation-region bounds and Jacobian.*

The interpolation region is defined by

$$\varOmega_p = \{(r,\phi):\ 1 < r < s,\ 0 \le \phi < 2\pi\}.$$

as derived in the previous section.

To obtain the expression of the Jacobian, note that the map $\boldsymbol{x} = \boldsymbol{P}_\eta^{-1/2}\boldsymbol{u}$ has constant determinant

$$\det(\boldsymbol{P}_\eta^{-1/2}) = \eta.$$

Thus

$$d\boldsymbol{x} = \det(\boldsymbol{P}_\eta^{-1/2})\, d\boldsymbol{u} = \eta\, d\boldsymbol{u}.$$

In polar coordinates, $d\boldsymbol{u} = r\, dr\, d\phi$, therefore

$$d\boldsymbol{x} = \eta\, r\, dr\, d\phi.$$

By definition,

$$E(\theta) = \int_{\Omega_p} \operatorname{tr}\big(\widetilde{\boldsymbol{F}}^T \widetilde{\boldsymbol{F}}\big)\, d\boldsymbol{x}.$$

Using $d\boldsymbol{x} = \eta\, r\, dr\, d\phi$ and $\Omega_p = \{1 < r < s,\ 0 \le \phi < 2\pi\}$, we obtain

$$E(\theta) = \eta \int_{r=1}^{s} \int_{\phi=0}^{2\pi} \operatorname{tr}\big(\widetilde{\boldsymbol{F}}(r,\phi)^T \widetilde{\boldsymbol{F}}(r,\phi)\big)\, r\, d\phi\, dr.$$

Define

$$\boldsymbol{G}(r,\phi) := \boldsymbol{\varDelta x}(r,\phi) \otimes \nabla_{\boldsymbol{x}} z(r,\phi) + (z(r,\phi) - s^2)\varDelta.$$

Since $z(r,\phi) = r^2$, this is

$$\boldsymbol{G}(r,\phi) = \boldsymbol{\varDelta x}(r,\phi) \otimes \nabla_{\boldsymbol{x}} z(r,\phi) + (r^2 - s^2)\varDelta.$$

Then

$$\widetilde{\boldsymbol{F}}(r,\phi) = \overline{\boldsymbol{F}} + \frac{1}{1-s^2}\boldsymbol{G}(r,\phi).$$

Compute

$$\begin{aligned}\widetilde{\boldsymbol{F}}^T \widetilde{\boldsymbol{F}} &= \left(\overline{\boldsymbol{F}} + \frac{1}{1-s^2}\boldsymbol{G}\right)^T \left(\overline{\boldsymbol{F}} + \frac{1}{1-s^2}\boldsymbol{G}\right) \\ &= \overline{\boldsymbol{F}}^T \overline{\boldsymbol{F}} + \frac{1}{1-s^2}(\overline{\boldsymbol{F}}^T \boldsymbol{G} + \boldsymbol{G}^T \overline{\boldsymbol{F}}) + \frac{1}{(1-s^2)^2}\boldsymbol{G}^T \boldsymbol{G}.\end{aligned}$$

Take the trace. Using $\operatorname{tr}(\overline{\boldsymbol{F}}^T \boldsymbol{G}) = \operatorname{tr}(\boldsymbol{G}^T \overline{\boldsymbol{F}})$,

$$\operatorname{tr}\big(\widetilde{\boldsymbol{F}}^T \widetilde{\boldsymbol{F}}\big) = \operatorname{tr}(\overline{\boldsymbol{F}}^T \overline{\boldsymbol{F}}) + \frac{2}{1-s^2}\operatorname{tr}(\overline{\boldsymbol{F}}^T \boldsymbol{G}) + \frac{1}{(1-s^2)^2}\operatorname{tr}(\boldsymbol{G}^T \boldsymbol{G}).$$

Since $\boldsymbol{G} = (\boldsymbol{\varDelta x} \otimes \nabla_{\boldsymbol{x}} z) + (r^2 - s^2)\varDelta$,

$$\operatorname{tr}(\overline{\boldsymbol{F}}^T \boldsymbol{G}) = \operatorname{tr}\big(\overline{\boldsymbol{F}}^T (\boldsymbol{\varDelta x} \otimes \nabla_{\boldsymbol{x}} z)\big) + (r^2 - s^2)\operatorname{tr}(\overline{\boldsymbol{F}}^T \varDelta).$$

Use the identity $\operatorname{tr}(\boldsymbol{A}^T(\boldsymbol{a} \otimes \boldsymbol{b})) = \boldsymbol{a}^T \boldsymbol{A} \boldsymbol{b}$ with $\boldsymbol{A} = \overline{\boldsymbol{F}}$, $\boldsymbol{a} = \nabla_{\boldsymbol{x}} z$, $\boldsymbol{b} = \boldsymbol{\varDelta x}$:

$$\operatorname{tr}\big(\overline{\boldsymbol{F}}^T (\boldsymbol{\varDelta x} \otimes \nabla_{\boldsymbol{x}} z)\big) = (\boldsymbol{\varDelta x})^T\, \overline{\boldsymbol{F}}(\nabla_{\boldsymbol{x}} z).$$

Therefore

$$\operatorname{tr}(\overline{\boldsymbol{F}}^T \boldsymbol{G}) = (\boldsymbol{\varDelta x})^T\, \overline{\boldsymbol{F}}(\nabla_{\boldsymbol{x}} z) + (r^2 - s^2)\operatorname{tr}(\overline{\boldsymbol{F}}^T \varDelta).$$

Expand:

$$\boldsymbol{G}^T \boldsymbol{G} = \big((\boldsymbol{\varDelta x} \otimes \nabla_{\boldsymbol{x}} z) + (r^2 - s^2)\varDelta\big)^T \big((\boldsymbol{\varDelta x} \otimes \nabla_{\boldsymbol{x}} z) + (r^2 - s^2)\varDelta\big).$$

Taking the trace and expanding:

$$\operatorname{tr}(\boldsymbol{G}^T \boldsymbol{G}) = \operatorname{tr}\big((\boldsymbol{\varDelta x} \otimes \nabla_{\boldsymbol{x}} z)^T (\boldsymbol{\varDelta x} \otimes \nabla_{\boldsymbol{x}} z)\big) + 2(r^2 - s^2)\operatorname{tr}\big((\boldsymbol{\varDelta x} \otimes \nabla_{\boldsymbol{x}} z)^T \varDelta\big) + (r^2 - s^2)^2 \operatorname{tr}(\varDelta^T \varDelta).$$

Use the identities:

$$\operatorname{tr}\big((\boldsymbol{a} \otimes \boldsymbol{b})^T (\boldsymbol{a} \otimes \boldsymbol{b})\big) = |\boldsymbol{a}|^2\, |\boldsymbol{b}|^2, \qquad \operatorname{tr}\big((\boldsymbol{a} \otimes \boldsymbol{b})^T \varDelta\big) = \boldsymbol{a}^T \varDelta\, \boldsymbol{b},$$

with $\boldsymbol{a} = \nabla_{\boldsymbol{x}} z$ and $\boldsymbol{b} = \boldsymbol{\varDelta x}$. Then

$$\operatorname{tr}(\boldsymbol{G}^T \boldsymbol{G}) = |\boldsymbol{\varDelta x}|^2\, |\nabla_{\boldsymbol{x}} z|^2 + 2(r^2 - s^2)(\boldsymbol{\varDelta x})^T \varDelta(\nabla_{\boldsymbol{x}} z) + (r^2 - s^2)^2 \operatorname{tr}(\varDelta^T \varDelta).$$

In the ellipse-aligned frame,

$$z(\boldsymbol{x}) = \frac{x_1^2}{\eta^2} + x_2^2.$$

Compute the gradient:

$$\nabla_{\boldsymbol{x}} z(\boldsymbol{x}) = \begin{pmatrix} \dfrac{\partial}{\partial x_1}\left(\dfrac{x_1^2}{\eta^2} + x_2^2\right) \\ \dfrac{\partial}{\partial x_2}\left(\dfrac{x_1^2}{\eta^2} + x_2^2\right) \end{pmatrix} = \begin{pmatrix} \dfrac{2x_1}{\eta^2} \\ 2x_2 \end{pmatrix}.$$

Substitute $x_1 = \eta\, r \cos\phi$ and $x_2 = r \sin\phi$:

$$\nabla_{\boldsymbol{x}} z(r,\phi) = \begin{pmatrix} \dfrac{2(\eta\, r\cos\phi)}{\eta^2} \\ 2(r\sin\phi) \end{pmatrix} = \begin{pmatrix} \dfrac{2\, r\cos\phi}{\eta} \\ 2\, r\sin\phi \end{pmatrix}.$$

Compute $\varDelta\boldsymbol{x}$:

$$\boldsymbol{x}(r,\phi) = \begin{pmatrix} \eta\, r\cos\phi \\ r\sin\phi \end{pmatrix}, \qquad \varDelta\boldsymbol{x}(r,\phi) = \varDelta\, \boldsymbol{x}(r,\phi) = \varDelta \begin{pmatrix} \eta\, r\cos\phi \\ r\sin\phi \end{pmatrix}.$$

Compute $|\nabla_{\boldsymbol{x}} z|^2$ explicitly:

$$\begin{aligned} |\nabla_{\boldsymbol{x}} z(r,\phi)|^2 &= \left(\frac{2\, r\cos\phi}{\eta}\right)^2 + (2r\sin\phi)^2 \\ &= 4r^2\left(\frac{\cos^2\phi}{\eta^2} + \sin^2\phi\right). \end{aligned}$$

Also note:

$$z(r,\phi) = r^2, \qquad z(r,\phi) - s^2 = r^2 - s^2.$$

*Assemble the full integrand.*
Insert the expansions for $\operatorname{tr}(\overline{\boldsymbol{F}}^T\boldsymbol{G})$ and $\operatorname{tr}(\boldsymbol{G}^T\boldsymbol{G})$ into the polar-coordinate integral of the energy $E(\theta)$ above to obtain:

$$E(\theta) = \eta \int_1^s \int_0^{2\pi} \left[\operatorname{tr}(\overline{\boldsymbol{F}}^T\overline{\boldsymbol{F}}) + \frac{2}{1-s^2}\operatorname{tr}(\overline{\boldsymbol{F}}^T\boldsymbol{G}) + \frac{1}{(1-s^2)^2}\operatorname{tr}(\boldsymbol{G}^T\boldsymbol{G})\right] r\, d\phi\, dr,$$

where

$$\boldsymbol{G}(r,\phi) = \varDelta\boldsymbol{x}(r,\phi) \otimes \nabla_{\boldsymbol{x}} z(r,\phi) + (r^2 - s^2)\varDelta.$$

Substitute $\operatorname{tr}(\overline{\boldsymbol{F}}^T\boldsymbol{G})$:

$$\operatorname{tr}(\overline{\boldsymbol{F}}^T\boldsymbol{G}) = (\varDelta\boldsymbol{x})^T\, \overline{\boldsymbol{F}}(\nabla_{\boldsymbol{x}} z) + (r^2 - s^2)\operatorname{tr}(\overline{\boldsymbol{F}}^T\varDelta),$$

and substitute $\operatorname{tr}(\boldsymbol{G}^T\boldsymbol{G})$:

$$\operatorname{tr}(\boldsymbol{G}^T\boldsymbol{G}) = |\varDelta\boldsymbol{x}|^2\, |\nabla_{\boldsymbol{x}} z|^2 + 2(r^2 - s^2)(\varDelta\boldsymbol{x})^T\varDelta(\nabla_{\boldsymbol{x}} z) + (r^2 - s^2)^2\operatorname{tr}(\varDelta^T\varDelta).$$

Hence the integrand becomes:

$$\begin{aligned} \operatorname{tr}\big(\widetilde{\boldsymbol{F}}^T\widetilde{\boldsymbol{F}}\big) &= \operatorname{tr}(\overline{\boldsymbol{F}}^T\overline{\boldsymbol{F}}) \\ &\quad + \frac{2}{1-s^2}\Big((\varDelta\boldsymbol{x})^T\, \overline{\boldsymbol{F}}(\nabla_{\boldsymbol{x}} z) + (r^2 - s^2)\operatorname{tr}(\overline{\boldsymbol{F}}^T\varDelta)\Big) \\ &\quad + \frac{1}{(1-s^2)^2}\Big(|\varDelta\boldsymbol{x}|^2\, |\nabla_{\boldsymbol{x}} z|^2 + 2(r^2 - s^2)(\varDelta\boldsymbol{x})^T\varDelta(\nabla_{\boldsymbol{x}} z) + (r^2 - s^2)^2\operatorname{tr}(\varDelta^T\varDelta)\Big). \end{aligned}$$

Therefore the full energy integral is:

$$\begin{aligned} E(\theta) = \eta \int_1^s \int_0^{2\pi} \Bigg[ &\operatorname{tr}(\overline{\boldsymbol{F}}^T\overline{\boldsymbol{F}}) \\ &+ \frac{2}{1-s^2}\Big((\varDelta\boldsymbol{x})^T\, \overline{\boldsymbol{F}}(\nabla_{\boldsymbol{x}} z) + (r^2 - s^2)\operatorname{tr}(\overline{\boldsymbol{F}}^T\varDelta)\Big) \\ &+ \frac{1}{(1-s^2)^2}\Big(|\varDelta\boldsymbol{x}|^2\, |\nabla_{\boldsymbol{x}} z|^2 + 2(r^2 - s^2)(\varDelta\boldsymbol{x})^T\varDelta(\nabla_{\boldsymbol{x}} z) + (r^2 - s^2)^2\operatorname{tr}(\varDelta^T\varDelta)\Big)\Bigg] r\, d\phi\, dr, \end{aligned}$$

with

$$\nabla_{\boldsymbol{x}} z(r,\phi) = \begin{pmatrix} \dfrac{2\, r\cos\phi}{\eta} \\ 2\, r\sin\phi \end{pmatrix}, \qquad \varDelta\boldsymbol{x}(r,\phi) = \varDelta \begin{pmatrix} \eta\, r\cos\phi \\ r\sin\phi \end{pmatrix}, \qquad 1 < r < s, \;\; 0 \le \phi < 2\pi.$$

# 4. Three-dimensional extension of the variational construction

The closed-form expression in the main text is specialized to a two-dimensional plane-strain problem because the inclusion rotation is then represented by a single angle. The variational construction itself does not rely on this planar restriction. For a three-dimensional ellipsoidal inclusion, one may define

$$z(\boldsymbol{x}) = \boldsymbol{x} \cdot \boldsymbol{R}_\alpha^T \boldsymbol{P} \boldsymbol{R}_\alpha \boldsymbol{x},$$

where $\boldsymbol{P}$ is a positive-definite shape tensor and $\boldsymbol{R}_\alpha \in \mathrm{SO}(3)$ sets the initial ellipsoid orientation. The inclusion, interpolation layer, and far-field matrix are again described by $z \leq 1$, $1 < z < s^2$, and $z \geq s^2$, respectively. The same interpolation ansatz applies with the rigid inclusion motion

$$\boldsymbol{y} = \boldsymbol{R}_0 \boldsymbol{x} + \boldsymbol{c}_0, \qquad \boldsymbol{R}_0 \in \mathrm{SO}(3),$$

and the optimal rigid rotation is obtained by minimizing the upper-bound energy over $\mathrm{SO}(3)$ and the translation. In contrast to the planar case, this generally produces a three-parameter nonlinear minimization problem rather than a single arctangent formula. This is the sense in which the main-text formula should be read as a compact 2D specialization of the general micromechanical interpolation approach.

# 5. Details of the finite-element implementation

The full-field finite-element calculations were carried out in FEniCSx and used only as numerical validation of the single-inclusion mechanical problem. The MIA formula is not fitted to these finite-element results.

- **Element type/order:** first-order Lagrange elements were used for the displacement field in both the matrix and inclusion.
- **Quadrature:** the weak form was integrated with a Gauss–Legendre quadrature rule of degree 8.
- **Boundary condition:** the outer boundary was subjected to affine displacement corresponding to the imposed average deformation gradient, equivalently $\boldsymbol{y} = \overline{\boldsymbol{F}}_0 \boldsymbol{x}$ on $\partial\Omega$.
- **Nonlinear solver:** Newton's method with a PETSc backtracking line-search algorithm was used, with relative tolerance $10^{-6}$ and absolute tolerance $10^{-7}$.
- **Penalty parameters:** near-incompressibility and inclusion rigidity were enforced using penalty terms with $\kappa = 10^6\mu$ and $\gamma = 10^7\mu$, respectively, where $\mu$ is the matrix shear modulus.
- **Rotation extraction:** the inclusion rotation was extracted by tracking the change in orientation of the major axis of the elliptical inclusion after deformation.
- **Mesh sensitivity:** the representative mesh used in the main comparisons contained 3542 elements with average element size approximately 0.067. The rotation response was compared against finer meshes with 13508 elements, average size approximately 0.034, and 146970 elements, average size approximately 0.01. The maximum absolute rotation difference between coarse mesh (3542 elements) relative the the finest (146970 elements) is $8\%$ while the fine mesh (13508 elements) relative to the finest is $3\%$.

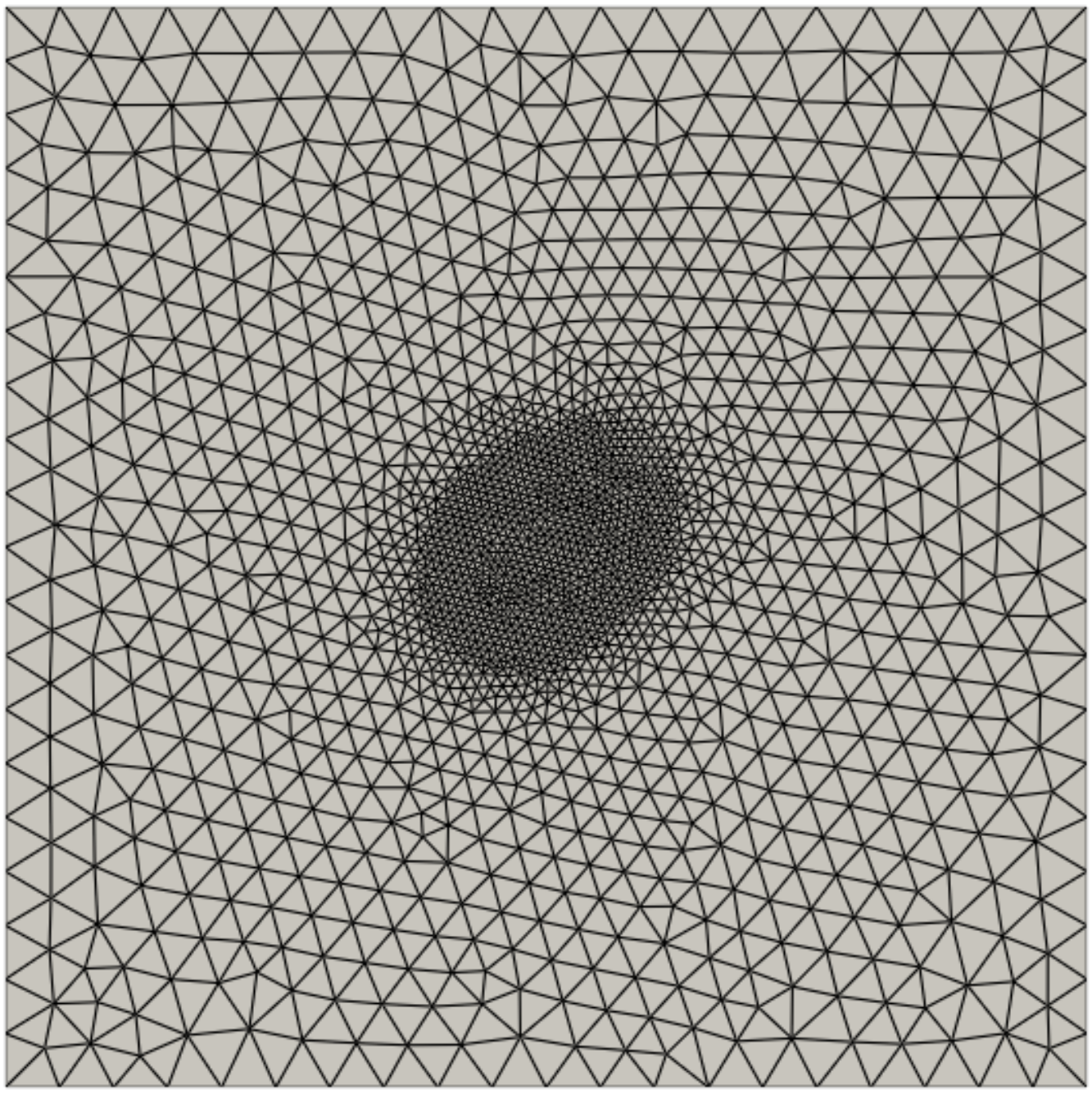

**Figure 1** Representative finite-element mesh containing 3542 first-order elements, with average element size approximately 0.067.

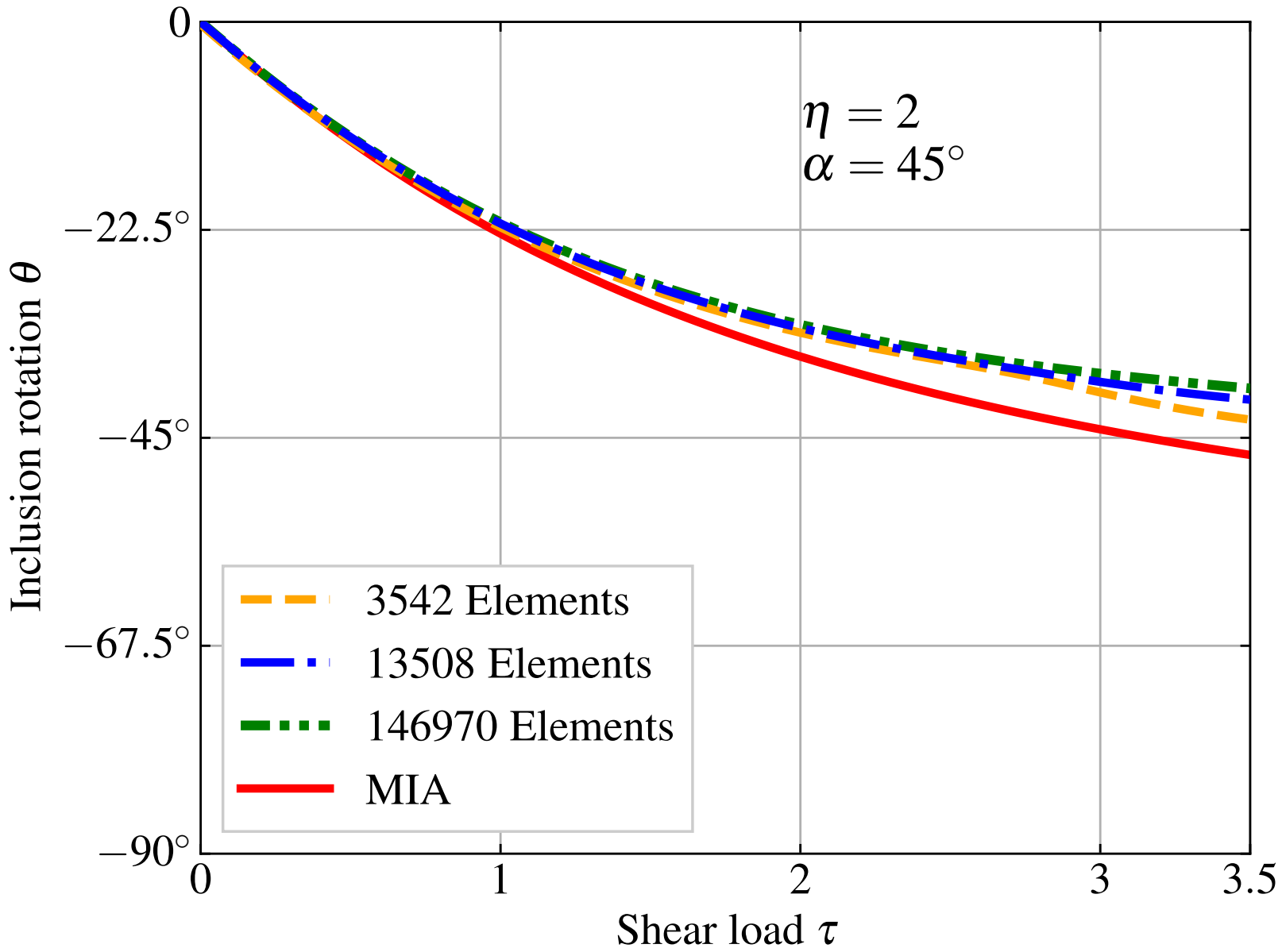


**Figure 2** Mesh-sensitivity comparison for inclusion rotation in the supplied simple-shear calculation with geometric orientation $\alpha_{\mathrm{g}} = 45^{\circ}$ and aspect ratio $\eta = 2$, using 3542, 13508, and 146970 elements with average element sizes approximately 0.067, 0.034, and 0.01, respectively.

We also varied the size of the computational cell containing one inclusion to assess finite-domain, or equivalently volume-fraction, sensitivity of the single-inclusion calculation. The supplied comparisons use geometric orientation $\alpha_{\mathrm{g}} = 60^{\circ}$ for shear test and $\alpha_{\mathrm{g}} = 45^{\circ}$ for tensile test, and aspect ratio $\eta = 2$.

Qualitatively, the shear curves remain comparatively close, whereas the tensile curves at the higher tested fractions ($10\%$ and $31\%$) visibly separate from MIA but relatively better than material mapping.

# 6. Isochoric projection

In this section, we seek to enforce incompressibility in 2D by a restricted trial family via $\boldsymbol{F} \mapsto \boldsymbol{F}/\sqrt{\det \boldsymbol{F}}$ so that $\det(\boldsymbol{F}) = 1$ pointwise by construction.

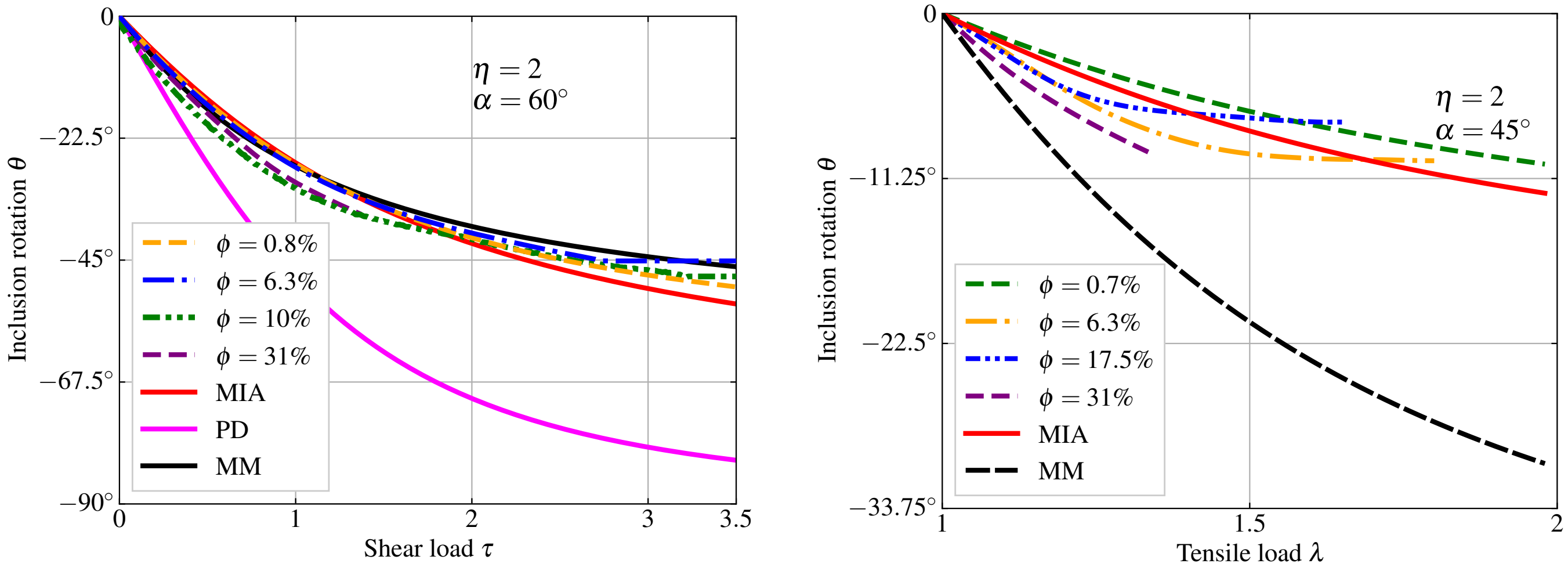


**Figure 3** Single-inclusion finite-domain sensitivity for (a) shear and (b) uniaxial tension. Each panel compares the MIA rotation with the finite-element result as the inclusion volume fraction $\phi$ is changed by varying the size of the one-inclusion computational cell.

## 6.1. Kinematics in the interpolation layer

We work in the ellipse-aligned frame ($\alpha = 0$) and use the rescaled polar coordinates from Section 2:

$$z = r^2, \qquad \boldsymbol{x} = \begin{pmatrix} \eta\, r\cos\phi \\ r\sin\phi \end{pmatrix}, \qquad \nabla_{\boldsymbol{x}} z = \begin{pmatrix} \dfrac{2\,r\cos\phi}{\eta} \\ 2\,r\sin\phi \end{pmatrix}, \qquad 1 < r < s,\ 0 \le \phi < 2\pi.$$

Let $\boldsymbol{R}(\theta)$ be the inclusion rotation and define $\boldsymbol{\varDelta} := \boldsymbol{R}(\theta) - \boldsymbol{F}$. The manuscript expression (paper Eq. (5)) can be written (with $\boldsymbol{c}_0 = \boldsymbol{0}$) as

$$\widetilde{\boldsymbol{F}}(r,\phi;\theta) = \boldsymbol{F} + \frac{\boldsymbol{\varDelta x} \otimes \nabla z + (r^2 - s^2)\boldsymbol{\varDelta}}{1 - s^2} \qquad \text{in } \varOmega_p. \tag{3}$$

The Jacobian is $d\boldsymbol{x} = \eta\, r\, dr\, d\phi$.

## 6.2. Isochoric projection

Define the projected (isochoric) deformation gradient

$$\widetilde{\boldsymbol{F}}_{\mathrm{iso}}(r,\phi;\theta) := \frac{\widetilde{\boldsymbol{F}}(r,\phi;\theta)}{\sqrt{\det \widetilde{\boldsymbol{F}}(r,\phi;\theta)}}. \tag{4}$$

Then $\det(\widetilde{\boldsymbol{F}}_{\mathrm{iso}}) \equiv 1$ pointwise (where $\det \widetilde{\boldsymbol{F}} > 0$). The (relevant) $\theta$–dependent part of the energy in $\varOmega_p$ becomes

$$E_{\mathrm{iso}}(\theta) = \frac{\mu\,\eta}{2} \int_1^s \int_0^{2\pi} \operatorname{tr}\Big((\widetilde{\boldsymbol{F}}_{\mathrm{iso}})^T \widetilde{\boldsymbol{F}}_{\mathrm{iso}}\Big)\, r\, d\phi\, dr = \frac{\mu\,\eta}{2} \int_1^s \int_0^{2\pi} \frac{\operatorname{tr}(\widetilde{\boldsymbol{F}}^T \widetilde{\boldsymbol{F}})}{\det(\widetilde{\boldsymbol{F}})}\, r\, d\phi\, dr, \tag{5}$$

since in 2D

$$\operatorname{tr}\Big((\boldsymbol{F}/\sqrt{\det \boldsymbol{F}})^T (\boldsymbol{F}/\sqrt{\det \boldsymbol{F}})\Big) = \frac{\operatorname{tr}(\boldsymbol{F}^T\boldsymbol{F})}{\det \boldsymbol{F}}.$$

(Additive constants independent of $\theta$ are omitted.)

## 6.3. Equilibrium equation for inclusion rotation

Differentiate (5) using

$$\frac{d}{d\theta}\operatorname{tr}(\widetilde{\boldsymbol{F}}^T\widetilde{\boldsymbol{F}}) = 2\,\widetilde{\boldsymbol{F}} : \dot{\widetilde{\boldsymbol{F}}}, \qquad \frac{d}{d\theta}\det(\widetilde{\boldsymbol{F}}) = \operatorname{cof}(\widetilde{\boldsymbol{F}}) : \dot{\widetilde{\boldsymbol{F}}},$$

where $\dot{(\cdot)} := \dfrac{d}{d\theta}(\cdot)$ and $\boldsymbol{A} : \boldsymbol{B} := \operatorname{tr}(\boldsymbol{A}^T\boldsymbol{B})$. Then

$$0 = \frac{dE_{\mathrm{iso}}}{d\theta} = \frac{\mu\,\eta}{2} \int_1^s \int_0^{2\pi} \left[ \frac{2\,\widetilde{\boldsymbol{F}} : \dot{\widetilde{\boldsymbol{F}}}}{\det(\widetilde{\boldsymbol{F}})} - \frac{\operatorname{tr}(\widetilde{\boldsymbol{F}}^T\widetilde{\boldsymbol{F}})}{\det(\widetilde{\boldsymbol{F}})^2} \left(\operatorname{cof}(\widetilde{\boldsymbol{F}}) : \dot{\widetilde{\boldsymbol{F}}}\right) \right] r\, d\phi\, dr. \tag{6}$$

### 6.4. Small-volumetric-deviation limit (reported rotation formula)

In regimes where the trial field is already *nearly* isochoric (so $\det(\widetilde{\boldsymbol{F}}) = 1 + O(\varepsilon)$), we can expand (5) to first order.

Assume the macroscopic deformation is incompressible ($\det \boldsymbol{F} = 1$) and write

$$\widetilde{\boldsymbol{F}} = \boldsymbol{F} + \varepsilon \boldsymbol{K}, \qquad \det(\widetilde{\boldsymbol{F}}) = 1 + \varepsilon \left(\operatorname{cof}(\boldsymbol{F}) : \boldsymbol{K}\right) + O(\varepsilon^2), \qquad \operatorname{tr}(\widetilde{\boldsymbol{F}}^T \widetilde{\boldsymbol{F}}) = \operatorname{tr}(\boldsymbol{F}^T \boldsymbol{F}) + 2\varepsilon(\boldsymbol{F} : \boldsymbol{K}) + O(\varepsilon^2).$$

Then

$$\frac{\operatorname{tr}(\widetilde{\boldsymbol{F}}^T \widetilde{\boldsymbol{F}})}{\det(\widetilde{\boldsymbol{F}})} = \operatorname{tr}(\boldsymbol{F}^T \boldsymbol{F}) + 2\varepsilon(\boldsymbol{F} : \boldsymbol{K}) - \varepsilon \operatorname{tr}(\boldsymbol{F}^T \boldsymbol{F}) \left(\operatorname{cof}(\boldsymbol{F}) : \boldsymbol{K}\right) + O(\varepsilon^2).$$

The projection therefore adds the corrective term $-\operatorname{tr}(\boldsymbol{F}^T \boldsymbol{F}) \left(\operatorname{cof}(\boldsymbol{F}) : \boldsymbol{K}\right)$ to the trace-only calculation. For the specific $\boldsymbol{K}$ induced by (3), the integrated stationarity condition is reported in the following form, linear in $\sin\theta$ and $\cos\theta$. The intermediate $(r, \phi)$ integration algebra is not yet included here, so this subsection should not yet be read as a complete derivation:

$$N(\eta, \alpha, \boldsymbol{F}) \cos\theta - D(\eta, \alpha, \boldsymbol{F}) \sin\theta = 0 \qquad \Longleftrightarrow \qquad \theta = \tan^{-1}\left(\frac{N}{D}\right). \tag{7}$$

With $\alpha$ denoting the geometric inclusion orientation in the notation of the main manuscript, the rotation formula used in the comparisons is

$$\theta^{\mathrm{MIA}} = \tan^{-1}\left(\frac{(F_{11} - F_{22})(\eta^4 - 1)\sin(2\alpha) - (F_{12} + F_{21})(\eta^4 - 1)\cos(2\alpha) + (F_{12} - F_{21})(\eta^4 + 6\eta^2 + 1)}{(F_{12} + F_{21})(\eta^4 - 1)\sin(2\alpha) + (F_{11} - F_{22})(\eta^4 - 1)\cos(2\alpha) + (F_{11} + F_{22})(\eta^4 + 6\eta^2 + 1)}\right). \tag{8}$$

Outside the nearly-isochoric regime, the exact equilibrium condition is (6). It generally does *not* reduce to a single arctangent formula.

# 7. Magneto-Mechanical Coupling in a Homogenized Setting

We use the hard-magnetic idealization in which the remanent magnetization is locked to the particle frame for applied fields below the coercive or reprogramming scale. We write the externally applied field as $\boldsymbol{H}^{\mathrm{app}} = |\boldsymbol{H}^{\mathrm{app}}|\boldsymbol{e}_1$. The magnetization in the deformed configuration is $\boldsymbol{m} = |\boldsymbol{m}| \begin{bmatrix} \cos\left(\theta^{\mathsf{MIA}}(\alpha, \lambda, \eta) + \alpha\right) \\ \sin\left(\theta^{\mathsf{MIA}}(\alpha, \lambda, \eta) + \alpha\right) \end{bmatrix}$. More generally, to distinguish geometry from magnetization, let $\alpha_{\mathrm{g}}$ denote the initial orientation of the inclusion major axis and let $\beta$ denote the initial remanent-magnetization direction, both measured from $\boldsymbol{e}_1$. For $\boldsymbol{H}^{\mathrm{app}} = |\boldsymbol{H}^{\mathrm{app}}|\boldsymbol{e}_1$, a particle rotation $\theta^{\mathrm{MIA}}(\alpha_{\mathrm{g}}, \lambda, \eta)$ gives

$$\boldsymbol{m} = |\boldsymbol{m}_0| \begin{bmatrix} \cos\big(\beta + \theta^{\mathrm{MIA}}(\alpha_{\mathrm{g}}, \lambda, \eta)\big) \\ \sin\big(\beta + \theta^{\mathrm{MIA}}(\alpha_{\mathrm{g}}, \lambda, \eta)\big) \end{bmatrix}. \tag{9}$$

The magneto-mechanical energy has the Zeeman contribution with a negative sign,

$$E(\lambda) = W_{\mathrm{mech}}(\lambda) - \mu_0 \boldsymbol{H}^{\mathrm{app}} \cdot \boldsymbol{m}. \tag{10}$$

This illustrative calculation assumes a spatially uniform externally applied field. Forces driven by gradients of $\boldsymbol{H}^{\mathrm{app}}$, as well as translation or navigation caused by such gradients, are not included.

For the aligned specialization used in the formulas below, the remanent direction and geometric major axis coincide: $\beta = \alpha_{\mathrm{g}} = \alpha$. The absolute magnetization angle is then $\alpha + \theta^{\mathrm{MIA}}$, while its loading-induced change is $\theta^{\mathrm{MIA}}$. Using $\cos(\theta + \alpha) = \cos\theta\cos\alpha - \sin\theta\sin\alpha$, the uniaxial-stretch specialization of (8) gives

$$\cos\theta^{\mathsf{MIA}} = \frac{\left(\eta^4 - 1\right)\left(\lambda - \lambda^{-1}\right)\cos(2\alpha) + \left(\eta^4 + 6\eta^2 + 1\right)\left(\lambda + \lambda^{-1}\right)}{\sqrt{\left(\left(\eta^4 - 1\right)\left(\lambda - \lambda^{-1}\right)\sin(2\alpha)\right)^2 + \left(\left(\eta^4 - 1\right)\left(\lambda - \lambda^{-1}\right)\cos(2\alpha) + \left(\eta^4 + 6\eta^2 + 1\right)\left(\lambda + \lambda^{-1}\right)\right)^2}}$$

$$\sin\theta^{\mathsf{MIA}} = \frac{(\eta^4 - 1)\left(\lambda - \lambda^{-1}\right)\sin(2\alpha)}{\sqrt{\left(\left(\eta^4 - 1\right)\left(\lambda - \lambda^{-1}\right)\sin(2\alpha)\right)^2 + \left(\left(\eta^4 - 1\right)\left(\lambda - \lambda^{-1}\right)\cos(2\alpha) + \left(\eta^4 + 6\eta^2 + 1\right)\left(\lambda + \lambda^{-1}\right)\right)^2}}$$

For the incompressible uniaxial stretch $\lambda$, the corresponding total energy is

$$E(\lambda) = \frac{\mu}{2}\left(\lambda^2 + \lambda^{-2} - 2\right) - \mu_0 |\boldsymbol{H}^{\mathrm{app}}||\boldsymbol{m}_0| \left(\cos\theta^{\mathsf{MIA}}(\alpha,\lambda,\eta)\cos\alpha - \sin\theta^{\mathsf{MIA}}(\alpha,\lambda,\eta)\sin\alpha\right). \tag{11}$$

Set $\lambda = 1+\epsilon$ and define the magnetic energy scale and the two aspect-ratio combinations

$$B := \mu_0|\boldsymbol{H}^{\mathrm{app}}|\,|\boldsymbol{m}_0|, \qquad k := \eta^4 - 1, \qquad q := \eta^4 + 6\eta^2 + 1. \tag{12}$$

The leading small-strain expansion consistent with the energy above is

$$E(\epsilon) = -B\cos\alpha + 2B\frac{k}{q}\sin^2\alpha\cos\alpha\,\epsilon + 2\mu\epsilon^2 + O(B\epsilon^2, \epsilon^3). \tag{13}$$

Minimization of this truncated energy therefore yields

$$\epsilon = -\frac{B}{2\mu}\frac{k}{q}\sin^2\alpha\cos\alpha + O\left(\left(\frac{B}{\mu}\right)^2\right). \tag{14}$$

The available field-driven rotation result is shown in Fig. 4. It is a separate numerical configuration from the small-field derivation above: the derivation uses an inclined aligned inclusion with $\boldsymbol{H}^{\mathrm{app}} \parallel \boldsymbol{e}_1$, while the plotted calculation uses $\alpha_{\mathrm{g}} = \beta = 0^\circ$, $\eta = 2$, and $\boldsymbol{H}^{\mathrm{app}} \parallel -\boldsymbol{e}_2$. In the aligned case $\beta = \alpha_{\mathrm{g}}$, the plotted particle rotation is also the change in remanent-magnetization direction; the absolute magnetization direction is $\beta + \theta^{\mathrm{MIA}}$. Equation $(14)$ is generically nonzero for an inclined anisotropic inclusion but vanishes in the symmetry cases encoded by $(\eta^4 - 1)\sin^2\alpha\cos\alpha = 0$.

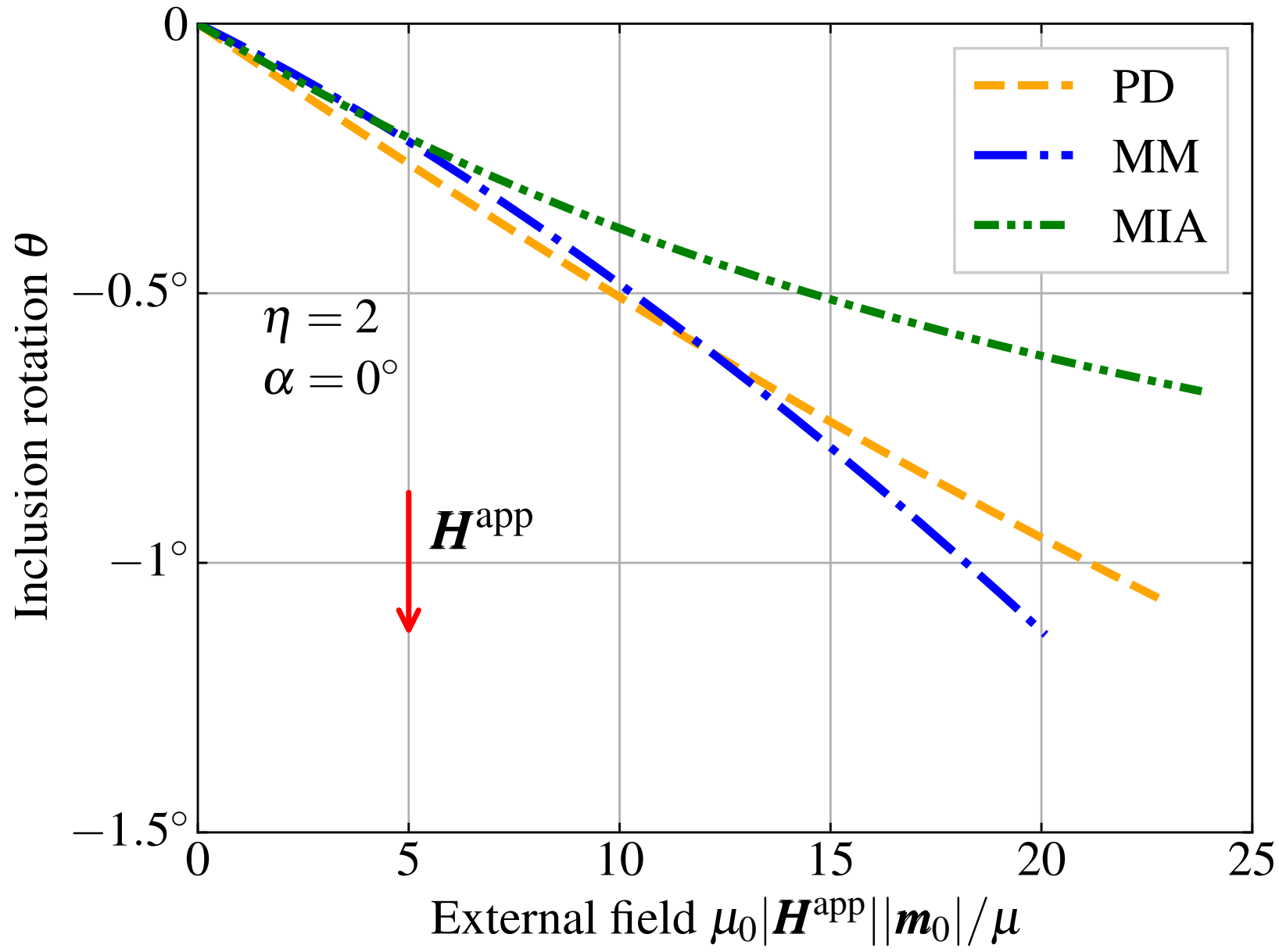


**Figure 4** Inclusion rotation versus the applied-field coordinate used in the supplied calculation, for aspect ratio $\eta = 2$, aligned initial directions $\alpha_{\mathrm{g}} = \beta = 0^\circ$, and an applied field directed along negative $e_2$. For this aligned case, the same ordinate gives the loading-induced change in magnetization direction.



\end{document}

%% file: definitions.tex
\usepackage{isomath}
\usepackage{amsmath}
\usepackage{amssymb}
\usepackage{amscd}
\usepackage{amsfonts}
\usepackage{pgf}
\usepackage{subfig}
\usepackage{natbib}
\usepackage{tabularx}

\newcommand{\figref}[1]{Fig.~\ref{#1}}

\newcommand{\tbfy}{{\widetilde{\bfy}}}

\newcommand{\tF}{\widetilde{\bfF}}
\newcommand{\barF}{\overline{\bfF}}
\newcommand{\OmInc}{\Omega^{\text{i}}}
\newcommand{\OmM}{\Omega^{\text{m}}}
\newcommand{\OmI}{\Omega^{\text{p}}}
\newcommand{\MUm}{\mu_{\text{m}}}

\newcommand{\la}{\lambda}

\DeclareMathOperator{\trace}{tr}

\newcommand{\bfb}{{\mathbold b}}
\newcommand{\bfc}{{\mathbold c}}

\newcommand{\bfe}{{\mathbold e}}

\newcommand{\bfm}{{\mathbold m}}

\newcommand{\bfx}{{\mathbold x}}
\newcommand{\bfy}{{\mathbold y}}

\newcommand{\bfB}{{\mathbold B}}

\newcommand{\bfF}{{\mathbold F}}

\newcommand{\bfH}{{\mathbold H}}
\newcommand{\bfI}{{\mathbold I}}

\newcommand{\bfM}{{\mathbold M}}

\newcommand{\bfP}{{\mathbold P}}

\newcommand{\bfR}{{\mathbold R}}

\newcommand{\bfU}{{\mathbold U}}